\documentclass[aps,prd,twocolumn]{revtex4}
\usepackage{amssymb}
\usepackage{amsfonts}
\usepackage{amsmath}
\usepackage{graphicx}
\usepackage{bm}
\usepackage{color}
\usepackage[dvips]{epsfig}
\usepackage[dvips]{graphicx}
\usepackage{float}
\usepackage{epstopdf}
\usepackage[colorlinks=true, pdfstartview=FitV, linkcolor=blue,citecolor=blue,urlcolor=blue, breaklinks=true]{hyperref}

\begin{document}

\title{Topological self-dual configurations in a Lorentz-violating gauged $%
O(3)$ sigma model}
\author{R. Casana}
\email{rodolfo.casana@gmail.com}
\author{C. F. Farias}
\email{cffarias@gmail.com}
\author{M. M. Ferreira, Jr.}
\email{manojr.ufma@gmail.com}
\affiliation{Departamento de F\'{\i}sica, Universidade Federal do Maranh\~{a}o,
65080-805 S\~{a}o Lu\'{\i}s, Maranh\~{a}o, Brazil.}

\begin{abstract}
We have studied the existence of topological Bogomol'nyi-Prasad-Sommerfield
or self-dual configurations
in a Lorentz-violating gauged $O(3)$ nonlinear sigma model, where $CPT$-even
Lorentz-violating (LV) terms were introduced in both the gauge and $\sigma $%
-field sectors. As happens in the usual gauged $\sigma $ model,
purely magnetic self-dual configurations are allowed, maintaining some
qualitative features of the standard ones. In a more involved configuration,
Lorentz violation provides new self-dual magnetic solutions carrying an
electric field but a null total electric charge. In both cases, the total
energy of the self-dual configurations turns out to be proportional to the
topological charge of the model and to the LV parameters introduced in the $%
\sigma $ sector. It is shown that the LV terms yield magnetic flux reversion
as well.
\end{abstract}

\maketitle

\section{Introduction}

In condensed matter physics, Abrikosov's description for type-II
superconductors \cite{ano} has led to an increasing interest in the study of
magnetic flux vortices that naturally stem from Ginzburg-Landau theory
\cite{ginz}. In field theory, stable vortex configurations were presented
for the first time by Nielsen and Olesen \cite{nielsen}, which showed that
electrically neutral vortices in the Maxwell-Higgs model correspond to the
Abrikosov ones. The existence of electrically charged vortex solutions was
verified in both the Chern-Simons-Higgs (CSH) \cite{jackiw1,jackiw2} and
Maxwell-Chern-Simons-Higgs (MCSH) \cite{lee} models. Vortices were also
investigated in the $O(3)$ model framework supplemented with an Abelian
gauge sector.

The nonlinear sigma $O(3)$ model in $\left( 1+2\right) $ dimensions \cite%
{polyakov} has become popular in field theory due to some remarkable
features and many possible applications to condensed matter physics \cite%
{CMP}. One feature which has attracted considerable attention is the fact it
provides topological stable solitonic solutions that are exactly integrable
in Bogomol'nyi limit \cite{bog}. These solutions can be described as a map
from a spherical surface that represents the two-dimensional physical space
to a spherical surface in the internal field space, being classified
according to the second homotopy group $\Pi _{2}\left( S_{2}\right) =Z.$
This model, however, presents a serious drawback: as the solutions are scale
invariant they can undergo arbitrary size changing over time with no energy
cost, preventing particle interpretation \cite{zakrzewski}. Some initial
ways to circumvent this difficulty were proposed, involving the
consideration of terms with a distinct number of derivative in relation to the
sigma model Lagrangian \cite{leese1}, or the construction of Q-lumps by
means of a particular potential in the $O(3)$ model \cite{leese2}. Another
interesting way to break the scale invariance consists of gauging the $U(1)$
subgroup and including a specific potential term in order to provide
self-dual solutions. This task was initially implemented by Schroers \cite%
{schroers}, with the dynamics of the gauge field being governed by a Maxwell
term, and providing topological soliton solutions with arbitrary magnetic
flux. The gauge dynamics can be also controlled by the Chern-Simons term
\cite{ghosh}, yielding topological and nontopological soliton solutions. In
both cases topological solitons are infinitely degenerate in a given
topological sector. Such a degenerescence can be lifted by choosing a
self-interacting potential with a symmetry breaking minima \cite%
{mukherjee1,mukherjee2}, which yields topological magnetic
Bogomol'nyi-Prasad-Sommerfield (BPS) vortices. This
potential introduces a new topology in which the infinite circle of physical
space is mapped in the equatorial circle in the internal space so that the
solitons are now classified by the first homotopy group $\Pi _{1}\left(
S_{1}\right) =Z$. Vortex configurations were also investigated in the
context of a modified gauged $O(3)$ model in which the gauge field is
nonminimally coupled to the $\sigma $ field \cite{almeida}.

The study of topological defects in several different theoretical frameworks
has been an issue of permanent interest in the recent years.
Solitons and vortex configurations have been addressed in field models
composed of generalizing dielectric functions with worthy new results \cite%
{bazeia}. Among these new investigations we can include there a search of
topological defects in field models endowed with Lorentz symmetry breaking
terms.

Lorentz symmetry violating field theories have been extensively investigated
since 1996, mainly in the framework of the standard model extension
\cite{kost97,coleman}.\textbf{\ }Such theoretical framework allows us to
examine the effects of Lorentz violation in physical systems, also involving
photon-fermion interactions \cite{Vertex}, bumblebee models \cite{bm},
fermion systems and radiative corrections \cite{fermion}, renormalization
 aspects \cite{new}, and the imposition of upper bounds on the
magnitude of the Lorentz-violating (LV) coefficients \cite{carroll,mewes1,altschul,klinkhamer1}.
LV theories are also connected to models containing higher-order derivative
terms \cite{HD} and higher-dimension operators \cite{Reyes}, and topological
aspects of\ physical systems \cite{anacleto,bakke}.

The formation of defects in field model with LV term was considered in some
situations, embracing solitons generated by scalar fields \cite{scalar},
Abelian monopoles \cite{barraz}, general defects engendered by tensor fields
\cite{seifert}, oscillon configurations \cite{correa1,correa3}.
Explicit BPS vortex solutions in Lorentz-violating scenarios were analyzed
in Refs. \cite{miller,casana1,sourrouille,belich,hott,Guillermo}.

Into the proposal of examining defects in new scenarios, this work aims at
elucidating how the structure of topological defects in a gauged $O(3)$
nonlinear sigma model is modified by Lorentz violation. More specifically,
we address the effects of introducing $CPT$-even and Lorentz-violating terms
both in the Abelian gauge sector and the $\sigma $ sector, as described in
Sec. II. The particular case in which the solutions are purely magnetic is
developed in Sec. III. In Sec. IV, the LV parameters are chosen to allow the
existence of magnetic configurations also carrying an electric field but a null
total electric charge. In both cases the BPS formalism is implemented,
yielding self-dual equations for the scalar and gauge fields. Some limit
cases on the LV parameters are discussed. We show that the energy and the
magnetic flux of the vortex solutions are proportional to the winding number
and depend explicitly on the LV parameters belonging to the $\sigma $%
 sector. In Sec. V, we present our conclusions and perspectives.

\section{The Lorentz-violating gauged $O(3)$ $\protect\sigma $ model}

The starting point is the $(1+2)$-dimensional Lagrangian density describing
the gauged $O(3)$ $\sigma $ model studied in Ref. \cite{mukherjee2},
enriched by $CPT$-even and Lorentz-violating terms,
\begin{eqnarray}
\mathcal{L} &=&\frac{1}{2}\left( D^{\mu }\vec{\phi}\right) \cdot \left(
D_{\mu }\vec{\phi}\right) +\frac{1}{2}{\left( k_{\phi \phi }\right) ^{\mu
\nu }\left( D_{\mu }\vec{\phi}\right) \cdot \left( D_{\nu }\vec{\phi}\right)
}  \notag \\[0.15cm]
&&-\frac{1}{4}F_{\mu \nu }F^{\mu \nu }-{\frac{1}{2}\kappa ^{\rho \alpha
}F_{\rho \sigma }F_{\alpha }{}^{\sigma }}-U,  \label{lg}
\end{eqnarray}%
where $\vec{\phi}=\left( \phi _{1},\phi _{2},\phi _{3}\right) $ is a triplet
of real scalar fields constituting a vector in the internal space with fixed
norm, $\vec{\phi}\cdot \vec{\phi}=1,$\ which describes an $O\left( 3\right) $
nonlinear $\sigma $ model. Such a scalar sector is coupled to the Maxwell
field, with $F_{\mu \nu }=\partial _{\mu }A_{\nu }-\partial _{\nu }A_{\mu }$
being the Maxwell tensor. The tensor $\left( k_{\phi \phi }\right) ^{\mu \nu
}$ is real and symmetric containing the LV and $CPT$-even parameters in the $%
\sigma $ sector, while $\kappa ^{\rho \alpha }$\ is a symmetric tensor which
encloses $CPT$-even and LV coefficients in the electromagnetic sector \cite%
{kost97,mewes1}. The potential $U$ describes some convenient interaction
producing BPS vortices.

The coupling between the gauge field and the triplet scalar field is given
by the minimal covariant derivative
\begin{equation}
D_{\mu }\vec{\phi}=\partial _{\mu }\vec{\phi}-A_{\mu }\hat{n}_{3}\times \vec{%
\phi},
\end{equation}%
with $\hat{n}_{3}$\ representing the 3-direction in the internal scalar
field space. The equation of motion for the gauge field reads
\begin{equation}
{\partial }_{\nu }F^{\nu \mu }+{\kappa ^{\nu \alpha }{\partial }_{\nu
}F_{\alpha }{}^{\mu }}-{\kappa ^{\mu \alpha }{\partial }_{\nu }F_{\alpha
}{}^{\nu }}=j^{\mu },
\end{equation}%
where
\begin{equation}
j^{\mu }=\left[ g^{\mu \nu }+{\left( k_{\phi \phi }\right) ^{\mu \nu }}%
\right] \hat{n}_{3}\cdot \left( \vec{\phi}\times D_{\nu }\vec{\phi}\right) ,
\end{equation}%
is the conserved current density that generalizes the one of Ref. \cite%
{mukherjee2}, that is, $j^{\mu }=\hat{n}_{3}\cdot \left( \vec{\phi}\times
D_{\nu }\vec{\phi}\right) .$ The equation of motion of the $\sigma $ field
is
\begin{eqnarray}
\left[ g^{\mu \nu }+{\left( k_{\phi \phi }\right) ^{\mu \nu }}\right] D_{\mu
}D_{\nu }\vec{\phi} &=&\left( \vec{\phi}\cdot \frac{\partial U}{\partial
\vec{\phi}}\right) \vec{\phi}-\frac{\partial U}{\partial \vec{\phi}} \\%
[0.15cm]
&&\hspace{-1.5cm}+\left[ g^{\mu \nu }+{\left( k_{\phi \phi }\right) ^{\mu
\nu }}\right] \left( \vec{\phi}\cdot D_{\mu }D_{\nu }\vec{\phi}\right) \vec{%
\phi}.  \notag
\end{eqnarray}

As we are interested in a solitonic solution in the static regime, we first write
the static Gauss law%
\begin{eqnarray}
L_{ij}{{\partial }_{i}\partial _{j}A_{0}}+{\kappa }_{0i}\epsilon _{ij}{{%
\partial }_{j}B} &=&{\left( k_{\phi \phi }\right) }_{0i}\hat{n}_{3}\cdot
\left( \vec{\phi}\times D_{i}\vec{\phi}\right) \\[0.15cm]
&&\hspace{-1.5cm}+\left[ 1+{\left( k_{\phi \phi }\right) }_{00}\right] \left[
\left( \phi _{1}\right) ^{2}+\left( \phi _{2}\right) ^{2}\right] A_{0},
\notag
\end{eqnarray}%
where we have introduced the symmetric matrix
\begin{equation}
L_{ij}=(1+{\kappa }_{00}){\delta }_{ij}-{\kappa }_{ij}.
\end{equation}

With the finality to attain self-dual configurations, we
have selected $\left( k_{\phi \phi }\right) _{0i}=0$. Such a choice allows us
to study two cases. The first one consists in choosing $\kappa _{0i}=0$, for
which the Gauss law is\textbf{\ }
\begin{equation}
L_{ij}{{\partial }_{i}\partial _{j}A_{0}}=\left[ 1+{\left( k_{\phi \phi
}\right) }_{00}\right] \left[ \left( \phi _{1}\right) ^{2}+\left( \phi
_{2}\right) ^{2}\right] A_{0}.  \label{Gauss_U0}
\end{equation}%
Because the condition $A_{0}=0$ satisfies identically the Gauss law, this
case describes purely magnetic solutions. The second case to consider is $%
\kappa _{0i}\neq 0$, this time the Gauss law becomes
\begin{equation}
L_{ij}{{\partial }_{i}\partial _{j}A_{0}}+{\kappa }_{0i}\epsilon _{ij}{{%
\partial }_{j}B}=[1+{(k_{\phi \phi })}_{00}][(\phi _{1})^{2}+(\phi
_{2})^{2}]A_{0},  \label{Gauss_U}
\end{equation}%
and the solutions also possess an electric field but the total electric charge
is zero.

By considering the condition ${\left( k_{\phi \phi }\right) }_{0i}{=0}$, the
static Amp\`ere law reads
\begin{eqnarray}
N_{ji}\partial _{i}B &=&-\kappa _{0i}\partial _{i}\partial _{j}A_{0}+\kappa
_{0j}\partial _{i}\partial _{i}A_{0} \\[0.15cm]
&&+\left[ \delta _{ji}-{\left( k_{\phi \phi }\right) }_{ji}\right] \hat{n}%
_{3}\cdot \left( \vec{\phi}\times D_{i}\vec{\phi}\right) ,  \notag
\end{eqnarray}%
where we have defined the antisymmetric matrix
\begin{equation}
N_{ji}=\epsilon _{ji}-\epsilon _{jm}\kappa _{mi}-\kappa _{jm}\epsilon _{mi}.
\end{equation}

In the following we study the two cases mentioned above. We first study the
purely magnetic solutions, and in the sequel, the ones carrying a magnetic and an
electric field.

\section{Purely magnetic self-dual configurations in a $CPT$-even and LV
gauged $O(3)$ $ \protect\sigma $ model}

The purely magnetic solutions are obtained by considering the conditions ${%
\left( k_{\phi \phi }\right) }_{0i}=0$, $\kappa _{0i}=0$, $A_{0}=0$, whose
energy is
\begin{equation}
E=\int d^{2}x\left( \frac{1}{2}\left( 1-\kappa _{ii}\right) B^{2}+\frac{1}{2}%
\tilde{D}_{k}\vec{\phi}\cdot \tilde{D}_{k}\vec{\phi}+U\right) ,  \label{ee0}
\end{equation}%
where $\kappa _{ii}=\kappa _{11}+\kappa _{22}$ and\ we have defined $\tilde{D%
}_{k}\vec{\phi}$ by%
\begin{eqnarray}
\tilde{D}_{k}\vec{\phi} &=&M_{kj}D_{j}\vec{\phi},  \label{mmx0} \\[0.06in]
\delta _{jk}-{\left( k_{\phi \phi }\right) _{jk}} &{=}&M_{ij}M_{ik},
\label{mmx1}
\end{eqnarray}%
where the coefficients $M_{ij}$ define the matrix $\mathbb{M}$ englobing the
spatial LV coefficients of the $\sigma $ sector. Before implementing the BPS
formalism, we introduce the identity%
\begin{eqnarray}
\frac{1}{2}\tilde{D}_{k}\vec{\phi}\cdot \tilde{D}_{k}\vec{\phi} &=&\frac{1}{4%
}\left( \tilde{D}_{j}\vec{\phi}\pm \epsilon _{jm}\vec{\phi}\times \tilde{D}%
_{m}\vec{\phi}\right) ^{2}  \label{trick} \\[0.15cm]
&&\mp \left( \det \mathbb{M}\right) \phi _{3}B\pm \left( \det \mathbb{M}%
\right) \epsilon _{ik}\partial _{i}\left( A_{k}\phi _{3}\right)  \notag \\%
[0.15cm]
&&\pm \left( \det \mathbb{M}\right) \vec{\phi}\cdot \left( \partial _{1}\vec{%
\phi}\times \partial _{2}\vec{\phi}\right) ,  \notag
\end{eqnarray}%
which allows us the express the energy (\ref{ee0})\ as
\begin{eqnarray}
E &=&\int d^{2}x\left\{ \frac{1}{4}\left( \tilde{D}_{j}\vec{\phi}\pm
\epsilon _{jm}\vec{\phi}\times \tilde{D}_{m}\vec{\phi}\right) ^{2}\right.
\label{ee1} \\
&&\hspace{0.5cm}+\frac{1}{2}\left( 1-\kappa _{ii}\right) \left( B\mp \sqrt{%
\frac{2U}{1-\kappa _{ii}}}\right) ^{2}  \notag \\[0.15cm]
&&\hspace{0.5cm}\pm \left( \det \mathbb{M}\right) \left[ \vec{\phi}\cdot
\left( \partial _{1}\vec{\phi}\times \partial _{2}\vec{\phi}\right)
+\epsilon _{ik}\partial _{i}\left( A_{k}\phi _{3}\right) \right]  \notag \\%
[0.15cm]
&&\hspace{0.5cm}\left. \pm B\left[ \sqrt{2\left( 1-\kappa _{ii}\right) U}%
-\left( \det \mathbb{M}\right) \phi _{3}\right] \right\} .  \notag
\end{eqnarray}%
The integration of the expression in the third row of Eq. (\ref{ee1}),%
\begin{equation}
T_{0}=\frac{\left( \det \mathbb{M}\right) }{4\pi }\!\int \!d^{2}x\!\left[
\vec{\phi}\cdot \left( \partial _{1}\vec{\phi}\times \partial _{2}\vec{\phi}%
\right) +\epsilon _{ik}\partial _{i}\left( A_{k}\phi _{3}\right) \right] ,
\label{tpc}
\end{equation}%
is the topological charge of the model, which depends on the Lorentz
violation introduced in the $\sigma $ sector - see the factor $\det \mathbb{M%
}$ and on the boundary conditions used to compute the integral. However,
the integrand keeps the same form of the one of Ref. \cite{mukherjee2}. In
Sec. \ref{vortex1}, we explicitly compute the associated topological
charge. We can also infer the existence of a conserved current,%
\begin{equation}
K_{\mu }=\frac{\left( \det \mathbb{M}\right) }{8\pi }\epsilon _{\mu \alpha
\beta }\left[ \vec{\phi}\cdot \left( D^{\alpha }\vec{\phi}\times D^{\beta }%
\vec{\phi}\right) +F^{\alpha \beta }\phi _{3}\right] ,
\end{equation}%
whose component $K_{0},$\ whenever integrated over the space, yields the
conserved topological charge (\ref{tpc}).

The fourth row of Eq. (\ref{ee1}) becomes null by choosing the potential,
\begin{equation}
U=\frac{\left( \det \mathbb{M}\right) ^{2}}{2\left( 1-\kappa _{ii}\right) }%
\phi _{3}^{2},
\end{equation}%
that is the one providing self-dual configurations. It corresponds to the
functional form of the one of Ref. \cite{mukherjee2}, $\phi _{3}^{2}/2,$
multiplied by LV terms, and presents the same minimal configurations, that
is,
\begin{equation}
\phi _{3}^{2}=0,\quad \phi _{1}^{2}+\phi _{2}^{2}=1.  \label{PM}
\end{equation}

Thus, the energy (\ref{ee1}) is written as
\begin{eqnarray}
E &=&\pm 4\pi T_{0}  \label{ee2} \\[0.2cm]
&&+\int d^{2}x\left\{ \frac{1}{4}\left( \tilde{D}_{j}\vec{\phi}\pm \epsilon
_{jm}\vec{\phi}\times \tilde{D}_{m}\vec{\phi}\right) ^{2}\right.  \notag \\
&&~\ \ \ \ \left. +\frac{1}{2}\left( 1-\kappa _{ii}\right) \left( B\mp \frac{%
\det \mathbb{M}}{1-\kappa _{ii}}\phi _{3}\right) ^{2}\right\} .  \notag
\end{eqnarray}

We finally notice that the energy (\ref{ee2}) has a lower bound
\begin{equation}
E\geq \pm 4\pi T_{0},  \label{E4Pi}
\end{equation}%
which is attained when the field configurations satisfy the self-dual or
BPS equations,
\begin{equation}
\tilde{D}_{j}\vec{\phi}\pm \epsilon _{jm}\vec{\phi}\times \tilde{D}_{m}\vec{%
\phi}=0,  \label{bbps_1}
\end{equation}%
\begin{equation}
B=\pm \frac{\det \mathbb{M}}{1-\kappa _{ii}}\phi _{3}.  \label{bbps_2}
\end{equation}%
Therefore, we have established the conditions that assure the existence of
purely magnetic self-dual configurations in a $CPT$-even and Lorentz-violating
 $O(3)$ $\sigma $ model. The sign $\pm $\ in Eq. (\ref{E4Pi}) indicates that
the topological charge $T_{0}$\ can be positive or negative, once the energy
is always positive.

For completeness, as happens in the Lorentz-invariant gauged model
of Refs. \cite{schroers,mukherjee2}, it is possible to show the existence of
an alternative form for the BPS equations (\ref{bbps_1}) and (\ref{bbps_2}) by
stereographically projecting the target space $S^{2}\ $into $C$\ $\cup
\left\{ \infty \right\} .$\ For it, we define the complex variable,
\begin{equation}
\omega =\frac{\phi _{1}+i\phi _{2}}{\left( 1+\phi _{3}\right) },
\end{equation}%
in terms of which the BPS equations are rewritten as
\begin{equation}
\tilde{\mathcal{D}}_{1}\omega =\mp i\tilde{\mathcal{D}}_{2}\omega ,
\label{BPSEG1}
\end{equation}%
\begin{equation}
F_{12}=\pm \frac{\left( \det \mathbb{M}\right) }{\left( 1-\kappa
_{ii}\right) }\frac{\left( 1-\left\vert \omega \right\vert ^{2}\right) }{%
\left( 1+\left\vert \omega \right\vert ^{2}\right) },  \label{BPSEG2}
\end{equation}%
with $\tilde{\mathcal{D}}_{k}=M_{kj}\left( \partial _{j}-iA_{j}\right) $.\ By
combining these equations, we obtain%
\begin{equation}
\left[ \delta _{jk}-{\left( k_{\phi \phi }\right) _{jk}}\right] \partial
_{j}\partial _{k}\ln \left\vert \omega \right\vert =\frac{\left( \det
\mathbb{M}\right) ^{2}}{1-\kappa _{ii}}\tanh \ln \left\vert \omega
\right\vert ,  \label{BPSEG3}
\end{equation}%
remembering that the matrices $\delta _{jk}-\left( k_{\phi \phi }\right)
_{jk}$\ and $M_{jk}$\ are related via Eq. (\ref{mmx1}). In the absence of
Lorentz-violation, $M_{kj}=\delta _{kj}$, $\kappa _{ii}=0$, and the BPS
equations (\ref{BPSEG1}) and (\ref{BPSEG2}) easily recover the ones of Ref. \cite%
{mukherjee2}, as expected.

Below we analyze a particular Ansatz describing the axially symmetric
vortices.

\subsection{Axially symmetrical purely magnetic self-dual solutions \label%
{vortex1}}

For the energy to be finite, the field $\vec{\phi}$\ should go to one of the
minimum configurations of the potential, stated in Eq. (\ref{PM}). This is
reached following a similar ansatz to the one introduced in\ Ref. \cite%
{mukherjee2} for axially symmetric vortices:
\begin{eqnarray}
\phi _{1} &=&\sin g(r)\cos \left( \frac{n}{\Lambda }\theta \right) ,~\phi
_{2}=\sin g(r)\sin \left( \frac{n}{\Lambda }\theta \right) ,  \notag \\%
[0.15cm]
\phi _{3} &=&\cos g(r),~A_{\theta }=-\frac{1}{r}\left[ a(r)-\frac{n}{\Lambda
}\right] ,  \label{ansatz}
\end{eqnarray}%
with the radial functions, $g$, $a,$\ being well behaved and satisfying the
boundary conditions,
\begin{eqnarray}
g(0) &=&0\,,~\ a(0)=\frac{n}{\Lambda },  \label{bc1} \\[0.15cm]
g(\infty ) &=&\frac{\pi }{2}\,,~\ a(\infty )=0,  \label{bc2}
\end{eqnarray}%
which are compatible with the vacuum configurations of the potential when $%
r\rightarrow \infty $. Here, $n$\ is the winding number, a non-null integer,
expressing the topological feature of the solutions. The boundary condition
of the vector potential is now modified by the presence of the constant $%
\Lambda $, defined in terms of the Lorentz-violating parameters belonging to
the $\sigma $ sector,\
\begin{equation}
\Lambda =\sqrt{\frac{1-\left( k_{\phi \phi }\right) _{\theta \theta }}{%
1-\left( k_{\phi \phi }\right) _{rr}}}.  \label{Lambda}
\end{equation}%
\textbf{\ }It is worthwhile to clarify the reason for introducing the LV
parameter $\Lambda $\ both in the ansatz (\ref{ansatz}) as in the boundary
condition $a(0)=n/\Lambda $. Its presence guarantees that when $r\rightarrow
0$, the profile $g(r)$\ is proportional to $r^{|n|}$\ [see Eqs. (\ref{zz1})
and (\ref{ggg2})], as happens in the usual vortex solutions. We
should also mention that the infinite circle (in coordinate space) is mapped
on the equatorial circle in the internal space $\vec{\phi}=(\phi _{1},\phi
_{2},0),$\ with $\phi _{1}^{2}+\phi _{2}^{2}=1.$\ The associated topological
solutions are not infinitely degenerated in each sector.

In the \textit{ansatz} (\ref{ansatz}), the magnetic field $B\ $reads,%
\begin{equation}
B(r)=-\frac{a^{\prime }}{r},
\end{equation}%
where ${}^{\prime }{}$\ stands for the radial derivative. In the same way,
the BPS\ equations (\ref{bbps_1}) and (\ref{bbps_2}) read
\begin{equation}
g^{\prime }=\pm \Lambda \frac{a}{r}\sin g,  \label{BPS3a}
\end{equation}%
\begin{equation}
B=-\frac{a^{\prime }}{r}=\pm \frac{\eta }{1-\kappa _{ii}}\cos g,
\label{BPS3b}
\end{equation}%
where $\eta $ is%
\begin{equation}
\eta =\det \mathbb{M}=\sqrt{\left[ 1-\left( k_{\phi \phi }\right) _{\theta
\theta }\right] \left[ 1-\left( k_{\phi \phi }\right) _{rr}\right] }.
\label{eta}
\end{equation}%
The expression (\ref{eta}) could lead us to interpret the matrix $\mathbb{M}$%
\ as being a diagonal one, which is not correct. It becomes clearer by
writing Eq. (\ref{mmx1}) in polar coordinates:
\begin{eqnarray}
\left( k_{\phi \phi }\right) _{rr} &=&1-M_{rr}^{2}-M_{\theta r}^{2},  \notag
\\
(k_{\phi \phi })_{r\theta } &=&(k_{\phi \phi })_{\theta r}=-M_{rr}M_{r\theta
}-M_{\theta \theta }M_{\theta r}, \\
\left( k_{\phi \phi }\right) _{\theta \theta } &=&1-M_{\theta \theta
}^{2}-M_{r\theta }^{2}.  \notag
\end{eqnarray}%
\ For the axially symmetric vortices, the condition $(k_{\phi \phi
})_{r\theta }=0$\ is a requirement for the BPS equation (\ref{bbps_1}) to
engender Eq. (\ref{BPS3a}). Consequently, the matrix $\left( k_{\phi \phi
}\right) $\ becomes diagonal, but the same does not occur with $\mathbb{M}$.
We should clarify that the fact of $(k_{\phi \phi })~$being diagonal in
polar coordinates does not imply that it will also be in Cartesian
coordinates.

Under the boundary conditions (\ref{bc1}) and (\ref{bc2}), the energy of the
self-dual solutions is%
\begin{equation}
E_{_{BPS}}=\pm 2\pi \frac{\eta }{\Lambda }n,
\end{equation}%
which, besides being proportional to the winding number, also depends
explicitly on the Lorentz violation factor $\eta /\Lambda ,$\ belonging to
the $\sigma $-sector. Here, positive (negative) sign is associated with
positive (negative) values of $n$.

The BPS energy density, $\varepsilon _{_{BPS}},$ which leads to the BPS
energy $\displaystyle E_{_{BPS}}=2\pi \int \!\!dr\,r\varepsilon _{_{BPS}}(r)$%
, is
\begin{equation}
\varepsilon _{_{BPS}}(r)=\left( 1-\kappa _{ii}\right) B^{2}+\Lambda \eta
\left( \frac{a}{r}\sin g\right) ^{2}.
\end{equation}%
It will be positive-definite whenever $\kappa _{ii}<1$ and $\Lambda \eta >0$.

By using the \emph{Ansatz} (\ref{ansatz}) and the boundary conditions (\ref%
{bc1})-(\ref{bc2}), we can compute the quantity $T_{0}$ in\ Eq. (\ref{tpc}),
\begin{eqnarray}
&&\left. \int d^{2}x~\vec{\phi}\cdot \left( \partial _{1}\vec{\phi}\times
\partial _{2}\vec{\phi}\right) =-2\pi \frac{n}{\Lambda }\int_{0}^{\infty
}dr~\left( \cos g\right) ^{\prime },\right.  \notag \\
&&\left. =-2\pi \frac{n}{\Lambda }\left[ \cos g(\infty )-\cos g(0)\right]
=2\pi \frac{n}{\Lambda },\right.
\end{eqnarray}%
\begin{equation}
\int d^{2}x~\epsilon _{ik}\partial _{i}\left( A_{k}\phi _{3}\right) =-2\pi
\int_{0}^{\infty }dr~\left[ \left( a-\frac{n}{\Lambda }\right) \cos g\right]
^{\prime }=0.  \notag
\end{equation}%
We thus show how the topological charge,
\begin{equation}
T_{0}=\frac{\eta }{2}\frac{n}{\Lambda },
\end{equation}%
is modified by the LV coefficients, recovering the usual charge, $T_{0}=n/2$%
, in the absence of Lorentz violation.

It is important to point out that the boundary conditions (\ref{bc1}) a (\ref%
{bc2}) provide solutions $g(r)$\ covering\ only the upper hemisphere of the
internal space. For solutions\ $g(r)$\ covering\ the lower hemisphere, we
use the following boundary conditions:
\begin{eqnarray}
g(0) &=&\pi \,,~\ a(0)=\frac{n}{\Lambda },  \label{lowerbc1} \\[0.15cm]
g(\infty ) &=&\frac{\pi }{2}\,,~\ a(\infty )=0.  \label{lowerbc2}
\end{eqnarray}%
A rapid analysis allows us to infer that the solutions corresponding to the
lower hemisphere can be obtained starting with the ones\ of the upper
hemisphere by doing the following correspondence
\begin{eqnarray}
g_{lower}\left( r\right) &=&\pi -g_{upper}\left( r\right) ,  \label{corresp1}
\\[0.15cm]
a_{lower}\left( r\right) &=&-a_{upper}\left( r\right) ,  \label{corresp2}
\end{eqnarray}%
where $g_{upper}\left( r\right) $\ and $a_{upper}\left( r\right) $\ are
solutions of the upper hemisphere for positive (negative) $n$\ and, $%
g_{lower}\left( r\right) $\ and $a_{lower}\left( r\right) $\ are solutions
of the lower hemisphere corresponding exactly to negative (positive) $n$.\
Consequently, for a given $n$\ the solutions in the upper and lower hemispheres
have opposite topological charges.

\subsubsection{Checking the behavior at boundaries}

By solving the BPS equations (\ref{BPS3a}) and (\ref{BPS3b}) near the origin,
one attains
\begin{eqnarray}
g(r) &\approx &G_{n}r^{n}+\ldots,  \label{zz1} \\[0.06in]
a(r) &\approx &{\frac{n}{\Lambda }}-\frac{\eta }{2\left( 1-\kappa
_{ii}\right) }r^{2}+\ldots,
\end{eqnarray}%
where $G_{n}$ is unique for a fixed $n$ and it is numerically computed. This
behavior is compatible with the boundary conditions (\ref{bc1}).

By solving the BPS equations (\ref{BPS3a}) and (\ref{BPS3b}) for $r\rightarrow
\infty $, the asymptotic behavior is
\begin{eqnarray}
g(r) &\approx &\frac{\pi }{2}-C_{_{\infty }}\frac{e^{-mr}}{\sqrt{r}}+\ldots,~ \\%
[0.15cm]
a(r) &\approx &\frac{mC_{_{\infty }}}{\Lambda }\sqrt{r}e^{-mr}+\ldots,
\end{eqnarray}%
with $C_{_{\infty }}$ being a numerically determined constant. Such behavior
is supported by the boundary conditions (\ref{bc2}). The constant $m$\ is
the mass of the self-dual bosons,\ given by
\begin{equation}
m=\sqrt{\frac{\eta \Lambda }{\left( 1-\kappa _{ii}\right) }}~.  \label{msc}
\end{equation}%
In the absence of Lorenz violation, the mass of the self-dual bosons is equal to
$1$. Considering LV corrections, the mass (\ref{msc}) can be larger or
smaller than 1, which corresponds to a narrower or wider vortex core,
respectively.

\subsection{Numerical analysis of the Lorentz-violating purely magnetic
self-dual vortices}

We first write the self-dual equations of the gauged $O(3)$ $\sigma $ model
in absence of Lorentz violation. These BPS equations can be directly
obtained from Eqs. (\ref{BPS3a}) and (\ref{BPS3b}) by setting $\Lambda =1$, $%
\eta =1$, and $\kappa _{ii}=0$, which reads
\begin{equation}
g^{\prime }=\pm \frac{a}{r}\sin g,  \label{SL_bp1}
\end{equation}%
\begin{equation}
B=-\frac{a^{\prime }}{r}=\pm \cos g.  \label{SL_bp2}
\end{equation}%
By comparing Eqs. (\ref{BPS3b}) and (\ref{SL_bp2}) it is easy to notice that
Lorentz violation can provide larger or smaller values for the magnetic
field amplitude at the origin. More detailed analysis of the
Lorentz-violating BPS solutions is performed by numerically solving the
differential equations (\ref{BPS3a}) and (\ref{BPS3b}). In particular, we
comment on the main aspects in which such solutions differ from the\ ones
obtained in the absence of Lorentz violation, described by Eqs. (\ref{SL_bp1}) and (%
\ref{SL_bp2}).

The BPS equations (\ref{BPS3a}) and (\ref{BPS3b}) provide a large family of
self-dual solutions, each one depending on the values of the
Lorentz-violating parameters. Among these many possibilities, we present the
solutions for $n=1$, $\kappa _{ii}=0.1$, $\eta =0.9999\Lambda $, and the
following values for $\Lambda $:
\begin{eqnarray}
\Lambda _{1} &=&0.5,~\Lambda _{2}=0.75,~\Lambda _{3}=1, \\[0.06in]
\Lambda _{4} &=&1.25,~\Lambda _{5}=1.5,~\Lambda _{6}=1.75.
\end{eqnarray}%
Such values of the Lorentz-violating parameters provide the following masses
for the self-dual bosons:
\begin{eqnarray}
m_{1} &=&0.52702,~m_{2}=0.79053,~m_{3}=1.05404, \\[0.06in]
m_{4} &=&1.31755,~m_{5}=1.58106,~m_{6}=1.84457,
\end{eqnarray}%
computed via Eq. (\ref{msc}), respectively.

\begin{figure}[]
\centerline{\includegraphics[width=8.0cm]{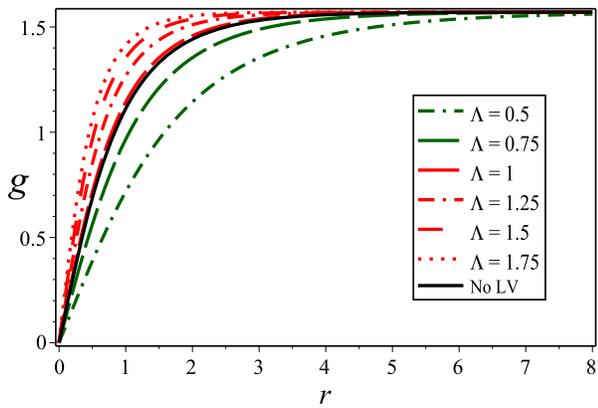}}
\caption{The profiles of the $\protect\sigma $ field, $g(r)$, for winding
number $n=1$. The black line is the profile in the absence of Lorentz violation.
The green lines depict the profiles of the self-dual solutions with masses $%
m<1$. The red lines represent self-dual solutions with masses $m>1$ . }
\label{S_BPS}
\end{figure}

\begin{figure}[]
\centerline{\includegraphics[width=8.0cm]{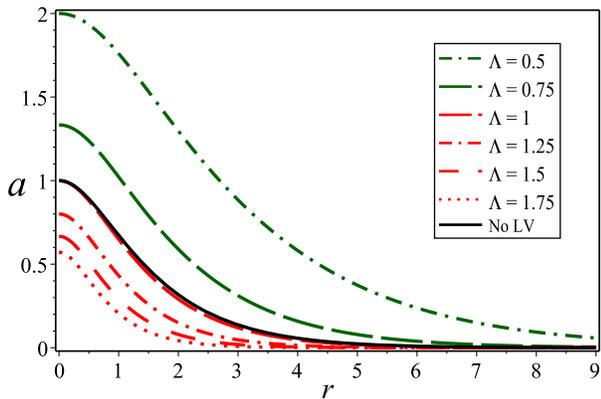}}
\caption{The profiles of the gauge field $a(r)$.}
\label{G_BPS}
\end{figure}

\begin{figure}[]
\centerline{\includegraphics[width=8.0cm]{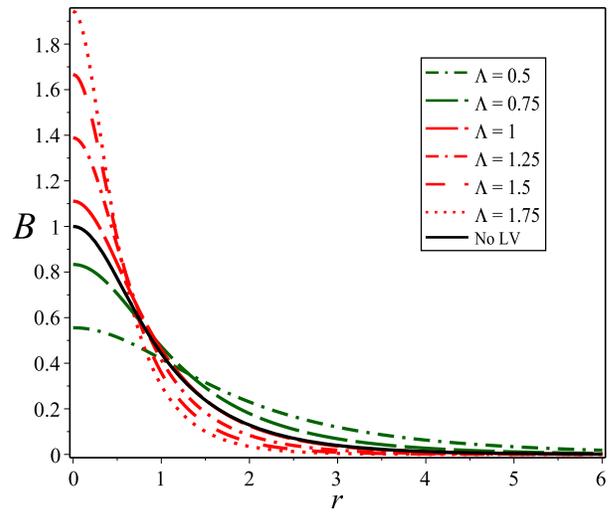}}
\caption{The profiles of the magnetic field $B(r)$.}
\label{B_BPS}
\end{figure}

\begin{figure}[]
\centerline{\includegraphics[width=8.0cm]{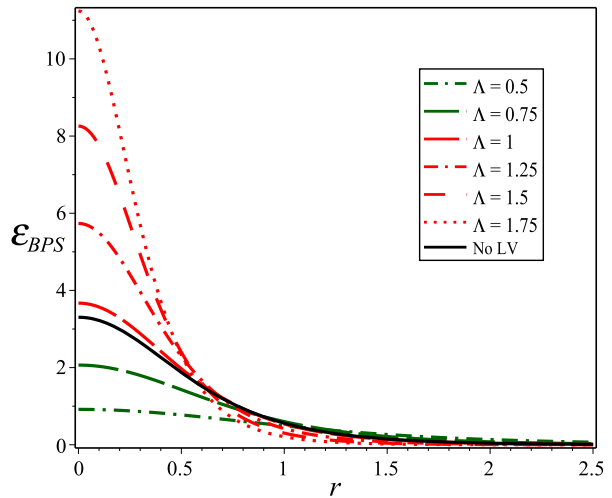}}
\caption{The profiles of the BPS energy density field $\protect\varepsilon %
_{_{BPS}}(r)$.}
\label{En_BPS}
\end{figure}

Figures \ref{S_BPS}--\ref{En_BPS} present some profiles (for the winding
number $n=1$) for the $\sigma $ field, gauge field, magnetic field, and BPS
energy density of the purely magnetic self-dual solutions. The black solid
line represents the BPS profiles in the absence of Lorentz violations. The green
lines represent the solutions with $\Lambda <1$ (or equivalently $m<1$),
while the red lines depict the ones with $\Lambda \geq 1$ (or equivalently $%
m>1$).

Figure \ref{S_BPS} depicts the numerical results obtained for the profiles
of the $\sigma $ field, showing that they turn out to be around the ones of the
model in the absence of Lorentz violation. These profiles become wider for $%
\Lambda <1$, otherwise, for $\Lambda \geq 1$, the profiles become
progressively narrower for increasing values of $\Lambda $, as expected.

Figure \ref{G_BPS} displays the profiles of the vector potential. As
happens with the $\sigma $-field profiles, they become wider for decreasing
values of $\Lambda <1$ and narrower for increasing values of $\Lambda \geq 1$%
. The novelty is the dependence of $a(0)$ in terms of $\Lambda ^{-1}$, which
is compatible with the boundary conditions imposed in Eq. (\ref{bc1}).

Figure \ref{B_BPS} depicts the magnetic field profiles. They are lumps
centered at the origin whose amplitudes are proportional to $\eta (1-\kappa
_{ii})^{-1}$. The LV parameters were fixed as $\kappa _{ii}=0.1$, $\eta
=0.9999\Lambda $, then for increasing values of $\Lambda $ (red lines),
amplitudes higher than 1 (black line, the amplitude in the absence of Lorentz
violation) and narrower profiles are obtained, otherwise, amplitudes smaller
than 1 and wider profiles are revealed for decreasing values of $\Lambda $
(green lines).

The profiles of the BPS energy density are shown in Fig. \ref{En_BPS}, which
also are lumps centered at origin like the ones of the magnetic field. Their
amplitudes at origin are given by $1.1109\Lambda ^{2}+0.9999\left(
G_{1}\right) ^{2}$, where $G_{1}$ is defined in (\ref{zz1}). Numerically, it
is shown that $G_{1}$ grows or diminishes with $\Lambda ,$ following its
behavior.

We remark that Lorentz violation works as a factor able to reduce or
increment the radial extension of the vortex core, the amplitude of the
magnetic field, and BPS energy density, keeping the topological character%
\textbf{\ }and enriching the diversity of purely magnetic self-dual
solutions of the gauged O(3) $\sigma $ model.

\section{Magnetic self-dual configurations carrying an electric field in a
$CPT$-even and LV gauged $O(3)$ $\protect\sigma $ model}

In this section, we describe the configurations possessing an electric field
but a null total electric charge, associated with the conditions $(k_{\phi
\phi })_{0i}=0$, $\kappa _{0i}\neq 0$, and the Gauss law (\ref{Gauss_U}). In
order to provide a correct description of magnetic self-dual configurations
carrying an electric field, we must modify the Lagrangian density (\ref{lg}) by
introducing a neutral scalar field $\Psi $,
\begin{eqnarray}
\mathcal{L} &=&-\frac{1}{4}F^{\mu \nu }F_{\mu \nu }-\frac{1}{2}\kappa _{\nu
\rho }F^{\mu \nu }F_{\mu }{}^{\rho } \\[0.15cm]
&&+\frac{1}{2}\left( D^{\mu }\vec{\phi}\right) \cdot \left( D_{\mu }\vec{\phi%
}\right) +\frac{1}{2}{\left( k_{\phi \phi }\right) ^{\mu \nu }\left( D_{\mu }%
\vec{\phi}\right) \cdot }\left( D_{\nu }\vec{\phi}\right)  \notag \\
&&+\frac{1}{2}\left( 1+\kappa _{00}\right) \partial _{\mu }\Psi \partial
^{\mu }\Psi +\frac{1}{2}\kappa ^{\mu \nu }\partial _{\mu }\Psi \partial
_{\nu }\Psi  \notag \\[0.15cm]
&&-\frac{1}{2}\left[ 1+{\left( k_{\phi \phi }\right) }_{00}\right] \left[
\left( \phi _{1}\right) ^{2}+\left( \phi _{2}\right) ^{2}\right] \Psi
^{2}-U\left( \phi _{3},\Psi \right) ,  \notag  \label{lcharged}
\end{eqnarray}%
with the kinetic term of this new field also affected by the
Lorentz-violating tensor $\kappa ^{\mu \nu }$. The introduction of a neutral
scalar field has the aim at providing a consistent description of the
self-dual configurations carrying an electric field in our model. A similar
situation was reported in the context of Maxwell-Chern-Simons gauged $O(3)$\
sigma model \cite{sigmaMCSH}. The introduction of a neutral scalar field is
a well established procedure for a consistent description of self-dual
configurations and it was first reported \cite{lee} in
Maxwell-Chern-Simons-Higgs models based in supersymmetric arguments. It was
also successfully implemented in other subsequent extensions \cite{bolog},
including Lorentz-violating Maxwell-Higgs models \cite{casana1,Guillermo}.
Furthermore, the introduction of the neutral field is a physical requirement
for the existence of a $N=2$\ extended supersymmetric version of the model
which supports charged solutions \cite{susycsh,almeida}. As was
shown by Witten and Olive \cite{WittenOlive}, the central charge of the
extended supersymmetric algebra is related to a topological quantum number
which is related to the existence of a Bogomol'nyi bound and vice-versa.

The energy of the static solutions carrying an electric field is
\begin{eqnarray}
E &=&\int d^{2}x\left\{ \frac{1}{2}\tilde{D}_{k}\vec{\phi}\cdot \tilde{D}_{k}%
\vec{\phi}+\frac{1}{2}\left( 1-\kappa _{ii}\right) B^{2}+U\right.
\label{xx1} \\
&&+\frac{1}{2}L_{ij}\left( \partial _{i}A_{0}\right) \left( \partial
_{j}A_{0}\right) +\frac{1}{2}L_{ij}\left( \partial _{i}\Psi \right) \left(
\partial _{j}\Psi \right)  \notag \\[0.15cm]
&&\left. +\frac{1}{2}[1+(k_{\phi \phi })_{00}][(\phi _{1})^{2}+(\phi
_{2})^{2}][(A_{0})^{2}+\Psi ^{2}]\right\} .  \notag
\end{eqnarray}%
By using the identity (\ref{trick}) and implementing the BPS formalism, the
energy becomes
\begin{eqnarray}
E &=&\int d^{2}x\left\{ \frac{1}{4}\left( \tilde{D}_{j}\vec{\phi}\pm
\epsilon _{jm}\vec{\phi}\times \tilde{D}_{m}\vec{\phi}\right) ^{2}\right.
\label{xx2} \\
&&+\frac{1}{2}\left( 1-\kappa _{ii}\right) \left( B\mp \sqrt{\frac{2U}{%
1-\kappa _{ii}}}\right) ^{2}  \notag \\
&&+\frac{1}{2}L_{ij}\left( \partial _{i}A_{0}\pm \partial _{i}\Psi \right)
\left( \partial _{j}A_{0}\pm \partial _{j}\Psi \right)  \notag \\
&&+\frac{1}{2}[1+(k_{\phi \phi })_{00}][(\phi _{1})^{2}+(\phi
_{2})^{2}][A_{0}\pm \Psi ]^{2}  \notag \\
&&\pm \left( \det \mathbb{M}\right) \left[ \vec{\phi}\cdot \left( \partial
_{1}\vec{\phi}\times \partial _{2}\vec{\phi}\right) +\epsilon _{ik}\partial
_{i}\left( A_{k}\phi _{3}\right) \right]  \notag \\[0.15cm]
&&\pm B\left[ \sqrt{2\left( 1-\kappa _{ii}\right) U}-\left( \det \mathbb{M}%
\right) \phi _{3}\right]  \notag \\[0.15cm]
&&\mp L_{ij}\left( \partial _{i}A_{0}\right) \left( \partial _{j}\Psi \right)
\notag \\[0.15cm]
&&\left. \frac{{}}{{}}\mp \lbrack 1+(k_{\phi \phi })_{00}][(\phi
_{1})^{2}+(\phi _{2})^{2}]A_{0}\Psi \right\}.  \notag
\end{eqnarray}%
With the Gauss law (\ref{Gauss_U}), the last term can be written as
\begin{equation}
L_{ij}\Psi \partial _{i}\partial _{j}A_{0}+\kappa _{0i}\epsilon
_{ij}\partial _{j}\left( \Psi B\right) -\kappa _{0i}\epsilon _{ij}B\partial
_{j}\Psi ,
\end{equation}%
which allows us to express the energy (\ref{xx2}) as
\begin{eqnarray}
E &=&\int d^{2}x\left\{ \frac{1}{4}\left( \tilde{D}_{j}\vec{\phi}\pm
\epsilon _{jm}\vec{\phi}\times \tilde{D}_{m}\vec{\phi}\right) ^{2}\right.
\label{xx3} \\
&&+\frac{1}{2}\left( 1-\kappa _{ii}\right) \left( B\mp \sqrt{\frac{2U}{%
1-\kappa _{ii}}}\right) ^{2}  \notag \\
&&+\frac{1}{2}L_{ij}\left( \partial _{i}A_{0}\pm \partial _{i}\Psi \right)
\left( \partial _{j}A_{0}\pm \partial _{j}\Psi \right)  \notag \\
&&+\frac{1}{2}\left[ 1+\left( k_{\phi \phi }\right) _{00}\right] \left[
\left( \phi _{1}\right) ^{2}+\left( \phi _{2}\right) ^{2}\right] \left[
A_{0}\pm \Psi \right] ^{2}  \notag \\[0.15cm]
&&\pm B\left[ \sqrt{2\left( 1-\kappa _{ii}\right) U}-\left( \det \mathbb{M}%
\right) \phi _{3}+\kappa _{0i}\epsilon _{ij}\partial _{j}\Psi \right]  \notag
\\
&&\pm \left( \det \mathbb{M}\right) \left[ \vec{\phi}\cdot \left( \partial
_{1}\vec{\phi}\times \partial _{2}\vec{\phi}\right) +\epsilon _{jk}\partial
_{j}\left( A_{k}\phi _{3}\right) \right]  \notag \\
&&\left. \frac{{}}{{}}\mp L_{ij}\partial _{j}\left( \Psi \partial
_{i}A_{0}\right) \mp \kappa _{0i}\epsilon _{ij}\partial _{j}\left( \Psi
B\right) \right\} .  \notag
\end{eqnarray}%
By requesting that the fifth row be null, we find\ the interaction potential $U$,
\begin{equation}
U=\frac{1}{2\left( 1-\kappa _{ii}\right) }\left[ \left( \det \mathbb{M}%
\right) \phi _{3}-\kappa _{0i}\epsilon _{ij}\partial _{j}\Psi \right] ^{2},
\end{equation}%
which involves the presence of derivative terms and is the correct one for
generating self-dual configurations.\ Potentials composed of derivative
terms have also been observed in other Lorentz-violating Maxwell-Higgs
models \cite{casana1,Guillermo}. Note that this potential is much more involved
than the one in Ref. \cite{mukherjee2}.

As happens in the purely magnetic case, the sixth row in Eq. (\ref%
{xx3}) provides the topological charge of the model; see Eq. (\ref{tpc}). By
considering the fields going to zero at infinity, the integration of the
terms in the seventh row gives a null contribution to the energy. Thus, the
energy of the solutions carrying an electric field\textbf{\ }becomes
\begin{eqnarray}
E &=&\pm 4\pi T_{0}  \label{xx4} \\
&&+\int d^{2}x\left\{ \frac{1}{4}\left( \tilde{D}_{j}\vec{\phi}\pm \epsilon
_{jm}\vec{\phi}\times \tilde{D}_{m}\vec{\phi}\right) ^{2}\right.  \notag \\%
[0.15cm]
&&+\frac{1}{2}\left( 1-\kappa _{ii}\right) \left( B\mp \frac{\left( \det
\mathbb{M}\right) \phi _{3}-\kappa _{0i}\epsilon _{ij}\partial _{j}\Psi }{%
1-\kappa _{ii}}\right) ^{2}  \notag \\[0.15cm]
&&+\frac{1}{2}L_{ij}\left( \partial _{i}A_{0}\pm \partial _{i}\Psi \right)
\left( \partial _{j}A_{0}\pm \partial _{j}\Psi \right)  \notag \\
&&\left. +\frac{1}{2}\left[ 1+\left( k_{\phi \phi }\right) _{00}\right] %
\left[ \left( \phi _{1}\right) ^{2}+\left( \phi _{2}\right) ^{2}\right] %
\left[ A_{0}\pm \Psi \right] ^{2}\right\} .  \notag
\end{eqnarray}%
Then, from Eq. (\ref{xx4}) we see that energy is bounded from below,
\begin{equation}
E\geq \pm 4\pi T_{0}.
\end{equation}%
This lower bound is attained with the fields satisfying the self-dual or BPS
equations,
\begin{equation}
\tilde{D}_{j}\vec{\phi}\pm \epsilon _{jm}\vec{\phi}\times \tilde{D}_{m}\vec{%
\phi}=0,
\end{equation}%
\begin{equation}
B=\pm \frac{\left( \det \mathbb{M}\right) \phi _{3}-\kappa _{0i}\epsilon
_{ij}\partial _{j}\Psi }{1-\kappa _{ii}},
\end{equation}%
\begin{equation}
\partial _{i}A_{0}\pm \partial _{i}\Psi =0,
\end{equation}%
\begin{equation}
A_{0}\pm \Psi =0.
\end{equation}

The condition $\Psi =\mp A_{0}$ saturates the two last equations. This way,
the solitonic solutions also carrying an electric field are described by the
BPS equations%
\begin{equation}
\tilde{D}_{j}\vec{\phi}\pm \epsilon _{jm}\vec{\phi}\times \tilde{D}_{m}\vec{%
\phi}=0,  \label{bq1}
\end{equation}%
\begin{equation}
B=\pm \frac{\left( \det \mathbb{M}\right) \phi _{3}}{1-\kappa _{ii}}+\frac{%
\kappa _{0i}\epsilon _{ij}\partial _{j}A_{0}}{1-\kappa _{ii}},  \label{bq2}
\end{equation}%
and the Gauss law%
\begin{equation}
L_{ij}{{\partial }_{i}\partial _{j}A_{0}}+{\kappa }_{0i}\epsilon _{ij}{{%
\partial }_{j}B}=[1+{(k_{\phi \phi })}_{00}][(\phi _{1})^{2}+(\phi
_{2})^{2}]A_{0}.  \label{bq3}
\end{equation}

Below we study the particular case of self-dual configurations, the axially
symmetric vortices.

\subsection{Axially symmetrical self-dual vortices carrying an electric field}

By considering the axially symmetrical \textit{ansatz} (\ref{ansatz}) and
\begin{equation}
A_{0}=A_{0}(r),
\end{equation}%
the projected BPS equations (\ref{bq1}) and (\ref{bq2}) are written as
\begin{equation}
g^{\prime }=\pm \Lambda \frac{a}{r}\sin g,  \label{eqt2a}
\end{equation}%
\begin{equation}
B=-\frac{a^{\prime }}{r}=\pm \frac{\eta \cos g}{1-\kappa _{ii}}-\frac{\kappa
_{0\theta }A_{0}^{\prime }}{1-\kappa _{ii}},  \label{eqt2b}
\end{equation}%
whereas the Gauss law (\ref{bq3}) reads%
\begin{equation}
\left( 1+\lambda _{r}\right) \frac{\left( rA_{0}^{\prime }\right) ^{\prime }%
}{r}-\kappa _{0\theta }\frac{\left( rB\right) ^{\prime }}{r}=\eta \Lambda
\Delta A_{0}\sin ^{2}g.  \label{eqt2c}
\end{equation}%
Here, we have introduced
\begin{eqnarray}
\lambda _{r} &=&\kappa _{00}-\kappa _{rr}, \\[0.15cm]
\Delta &=&\frac{1+{(k_{\phi \phi })}_{00}}{\eta \Lambda },  \label{Delta}
\end{eqnarray}%
and the constants $\Lambda $ and $\eta $ are defined in Eqs. (\ref{Lambda})
and (\ref{eta}), respectively.

The functions $g(r)$, $a(r)$ fulfill
the same boundary conditions introduced in Eqs. (\ref{bc1}) and (\ref{bc2}) and
the function $A_{0}(r)$ satisfies the boundary conditions (\ref%
{bc3})-(\ref{bc4}), as will be shown explicitly in the manuscript.

Although presenting an electric field, the self-dual configurations described
by Eqs. (\ref{eqt2a})-(\ref{eqt2c}) possess a null total electric charge. This
can be demonstrated easily by using the Gauss law (\ref{eqt2c}) where the
right-hand side defines the electric charge density $\rho =A_{0}\sin ^{2}g$,
whose integration provides the total electric charge of the self-dual
configuration,
\begin{equation}
Q=2\pi \int_{0}^{\infty }r\rho (r)dr.
\end{equation}%
By using the boundary conditions described in the previous paragraph, the
integration of the Gauss law provides a null electric charge, that is,\textbf{%
\ }%
\begin{equation}
Q=0.
\end{equation}

The BPS energy density is given by
\begin{eqnarray}
\varepsilon _{_{BPS}}(r) &=&\left( 1-\kappa _{ii}\right) B^{2}+\eta \Lambda
\left( \frac{a\sin g}{r}\right) ^{2}  \label{energy_BPS2} \\
&&+\eta \Lambda \Delta \left( A_{0}\sin g\right) ^{2}+\left( 1+\lambda
_{r}\right) \left( A_{0}^{\prime }\right) ^{2},  \notag
\end{eqnarray}%
and is defined positive providing that
\begin{equation}
\kappa _{ii}<1,~\Delta >0,~\lambda _{r}>-1.
\end{equation}

As in the purely magnetic case, the solutions $g(r)$\ obtained from\ Eqs. (%
\ref{eqt2a})-(\ref{eqt2c}) fulfilling the boundary conditions (\ref{bc1}), (%
\ref{bc2}), (\ref{bc3}), (\ref{bc4}) only cover the upper hemisphere of the
internal space. The solutions $g(r)$\ covering the\ lower hemisphere are
obtained by solving the Eqs. (\ref{eqt2a})-(\ref{eqt2c}) with the boundary
conditions (\ref{lowerbc1}) and (\ref{lowerbc2}). Consequently, the solutions\
can be obtained starting from the first ones making the correspondence (\ref%
{corresp1}) and (\ref{corresp2}) and
\begin{equation}
\left( A_{0}\right) _{lower}(r)=-\left( A_{0}\right) _{upper}(r),
\end{equation}%
where $\left( A_{0}\right) _{upper}(r)$\ represents solutions of the upper
hemisphere for positive (negative) $n$\ and \ $\left( A_{0}\right)
_{lower}(r)$\ provides solutions of the lower hemisphere\ corresponding to
negative (positive) $n$, respectively.\ Consequently, as happens in
the purely magnetic case, for a given $n$\ the solutions obtained for the
upper and lower hemispheres have opposite topological charge.

\subsection{Behavior of the profiles at boundaries}

By solving Eqs. (\ref{eqt2a})-(\ref{eqt2c}) near to the origin, we obtain
\begin{eqnarray}
g(r) &\approx &G_{n}r^{n}+\ldots,  \label{ggg2} \\[0.08in]
a(r) &\approx &{\frac{n}{\Lambda }}-\frac{1}{2}\frac{\eta \left( 1+\lambda
_{r}\right) }{\kappa _{0\theta }^{2}+\left( 1-\kappa _{ii}\right) \left(
1+\lambda _{r}\right) }r^{2}+\ldots,  \label{BH_a0} \\[0.08in]
A_{0}(r) &\approx &A_{0}(0)+\frac{\eta \kappa _{0\theta }}{\kappa _{0\theta
}^{2}+\left( 1-\kappa _{ii}\right) \left( 1+\lambda _{r}\right) }r+\ldots,\quad
\end{eqnarray}%
where $A_{0}(0)$ is determined numerically for every $n$. The last equation
allows to set explicitly the boundary condition for $A_{0}$ at the origin%
\begin{equation}
A_{0}^{\prime }(0)=\frac{\eta \kappa _{0\theta }}{\kappa _{0\theta
}^{2}+\left( 1-\kappa _{ii}\right) \left( 1+\lambda _{r}\right) }.
\label{bc3}
\end{equation}

By solving the BPS equations when $r\rightarrow \infty $, we attain
\begin{eqnarray}
g(r) &\approx &\frac{\pi }{2}-C_{_{\infty }}\frac{e^{-mr}}{\sqrt{r}}+...~,~~
\\[0.06in]
a(r) &\approx &\frac{mC_{_{\infty }}}{\Lambda }\sqrt{r}e^{-mr}+...~,~\ \  \\
A_{0}(r) &\approx &\frac{\left( 1-\kappa _{ii}\right) m^{2}-\eta \Lambda }{%
\Lambda \kappa _{0\theta }m}C_{_{\infty }}\frac{e^{-mr}}{\sqrt{r}}+...,
\label{bc-3a}
\end{eqnarray}
where $C_{_{\infty }}$ is a positive constant numerically determined. The
mass of the self-dual bosons, $m$, is given by
\begin{equation}
m=\sqrt{\frac{\eta \Lambda \left( \beta _{1}\pm \beta _{2}\right) }{2\left[
\kappa _{0\theta }^{2}+\left( 1-\kappa _{ii}\right) \left( 1+\lambda
_{r}\right) \right] }},  \label{betax}
\end{equation}%
with $\beta _{1}$ and $\beta _{2}$ being positive real numbers given by
\begin{eqnarray}
\beta _{1} &=&\left( 1+\lambda _{r}\right) +\Delta \left( 1-\kappa
_{ii}\right) ~,~~ \\
\beta _{2} &=&\sqrt{\left[ \left( 1+\lambda _{r}\right) -\Delta \left(
1-\kappa _{ii}\right) \right] ^{2}-4\Delta \kappa _{0\theta }^{2}}.
\end{eqnarray}

For $\beta _{2}$, the condition $\left[ \left( 1+\lambda _{r}\right) -\Delta
\left( 1-\kappa _{ii}\right) \right] ^{2}\geq 4\Delta \kappa _{0\theta
}^{2}, $ must be satisfied. The signs in Eq. (\ref{betax})\ will be used as
follows: $+\left( -\right) $ for $\left( 1+\lambda _{r}\right) -\Delta
\left( 1-\kappa _{ii}\right) >0\left( <0\right) $.

From Eq. (\ref{bc-3a}), we obtain the boundary condition for $A_{0}(r)$\
when $r\rightarrow \infty $,%
\begin{equation}
A_{0}\left( \infty \right) =0.  \label{bc4}
\end{equation}

To finish, we present some limited cases on the LV parameters. When $\kappa
_{0\theta }=0$, Eq. (\ref{betax}) recovers the mass scale of uncharged BPS
vortices:
\begin{equation}
m=\sqrt{\frac{\eta \Lambda }{1-\kappa _{ii}}}.
\end{equation}

On the other hand, when $\beta _{2}=0$, the parity-odd coefficient can be
expressed in terms of parity-even ones,
\begin{equation}
\kappa _{0\theta }=\pm \frac{\left\vert 1+\lambda _{r}-\Delta \left(
1-\kappa _{ii}\right) \right\vert }{2\sqrt{\Delta }},
\end{equation}%
and the mass scale becomes
\begin{equation}
m=\ \sqrt{\frac{2\eta \Lambda \Delta }{\left( 1+\lambda _{r}\right) +\Delta
\left( 1-\kappa _{ii}\right) }}.  \label{mxx}
\end{equation}

Another interesting possibility is to set $\kappa _{ii}=0$ and $\lambda _{r}=0$
in Eq.(\ref{betax}), i.e., we can consider null the LV\ parity-even
electromagnetic coefficients, getting
\begin{equation}
m=\left( \eta \Lambda \right) ^{1/2}\sqrt{\frac{1+\Delta \pm \sqrt{\left(
1-\Delta \right) ^{2}-4\Delta \kappa _{0\theta }^{2}}}{2\left( 1+\kappa
_{0\theta }^{2}\right) }}\ ,  \label{betax3}
\end{equation}%
with signal $+\left( -\right) $ for $1-\Delta >0\left( <0\right) $.
This situation also provides Abrikosov-Nielsen-Olesen-like
vortices whenever the condition,
\begin{equation}
\left( 1-\Delta \right) ^{2}\geq 4\Delta \kappa _{0\theta }^{2},
\end{equation}%
is satisfied. We remark that in the absence of LV in $\sigma $ sector ($%
\Delta =\eta =\Lambda =1$) it is impossible to obtain
Abrikosov-Nielsen-Olesen-like vortices when $\kappa _{ii}$\ and $\lambda
_{r} $ are null because the mass scale (\ref{betax3}) becomes a complex
number.

\subsection{Numerical analysis: A \textquotedblleft charged" vortex
configuration}

\begin{figure}[]
\centering{\includegraphics[width=8.0cm]{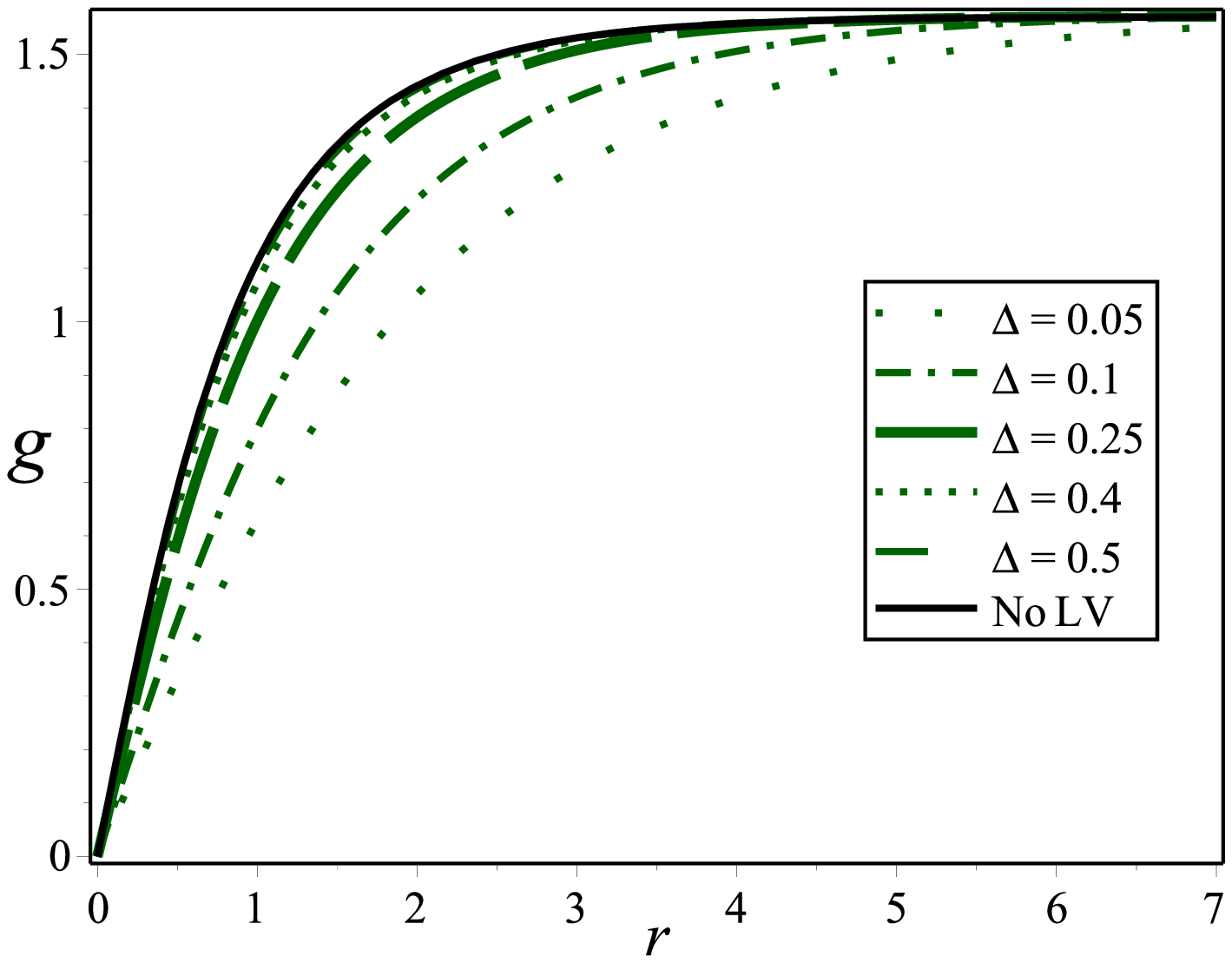} %
\includegraphics[width=8.0cm]{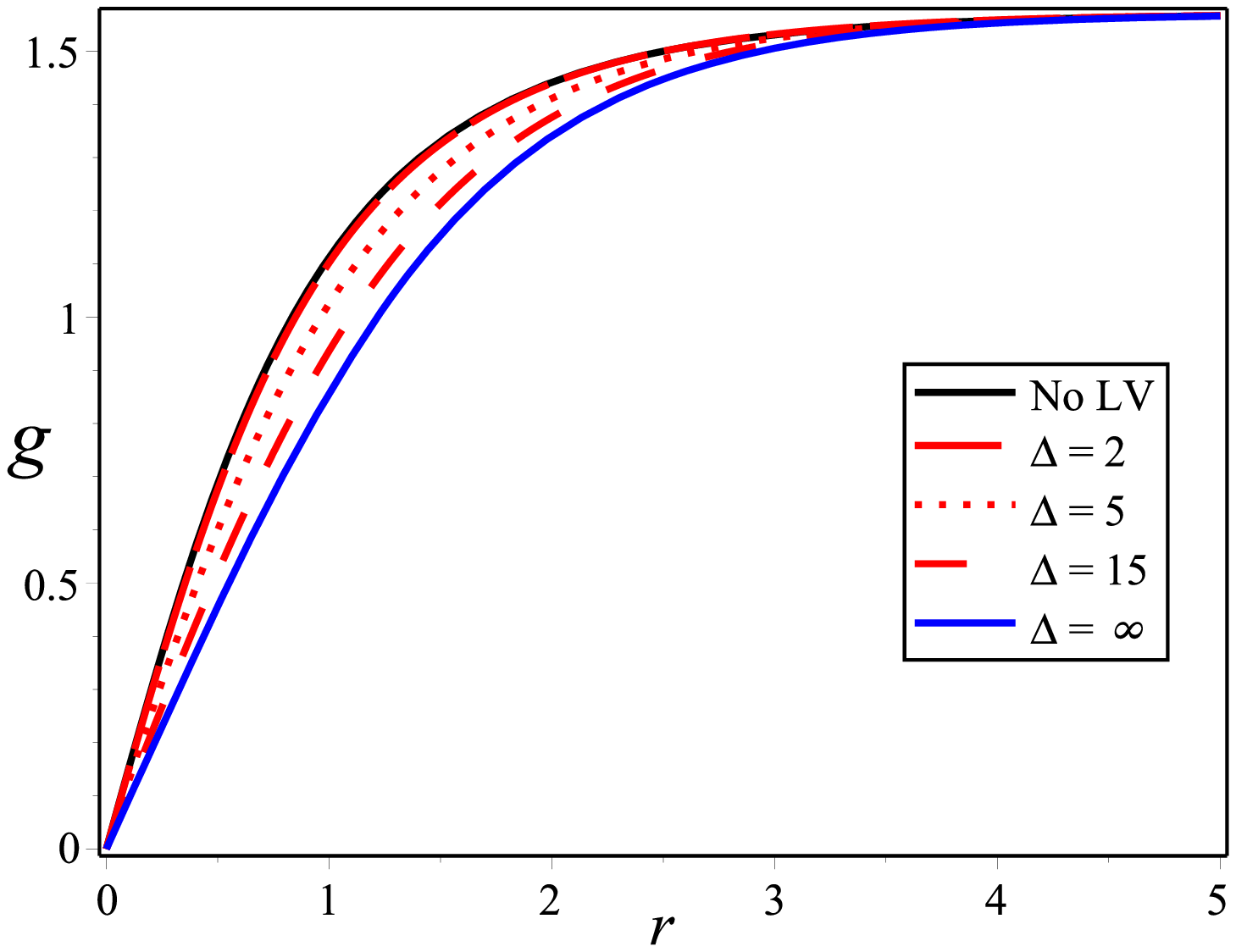}}
\caption{The profiles of the $\protect\sigma $ field, $g(r),$ for winding
number $n=1$. The upper figure represents the profiles for $0<\Delta \leq
0.5 $ and the lower figure represents the profiles for $\Delta \geq 2$. The
solid black line represents the profiles in absence of Lorentz violation.
The blue line is the profile for $\Delta \rightarrow \infty $.}
\label{HIGGS}
\end{figure}

\begin{figure}[]
\centering{\includegraphics[width=8.0cm]{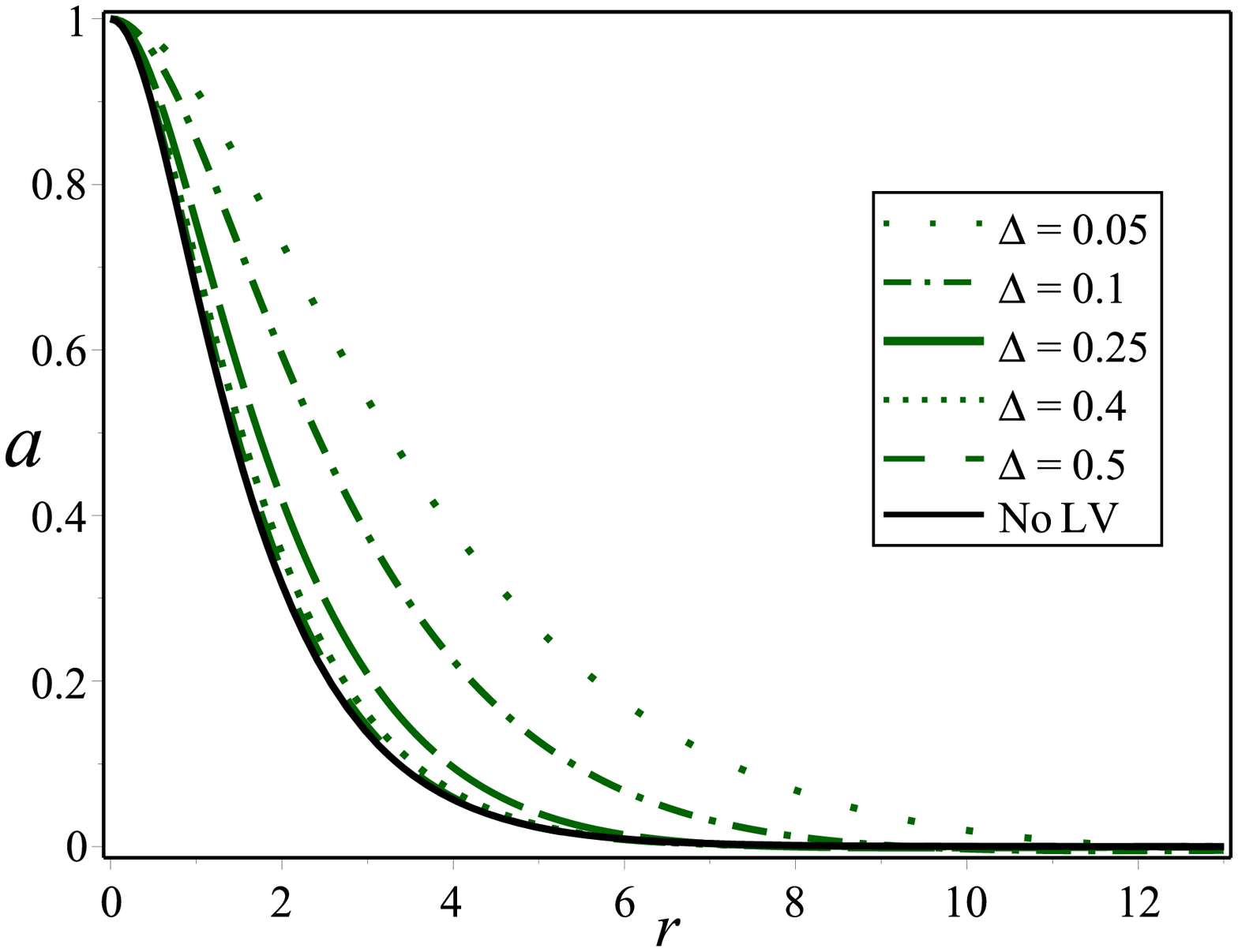} %
\includegraphics[width=8.0cm]{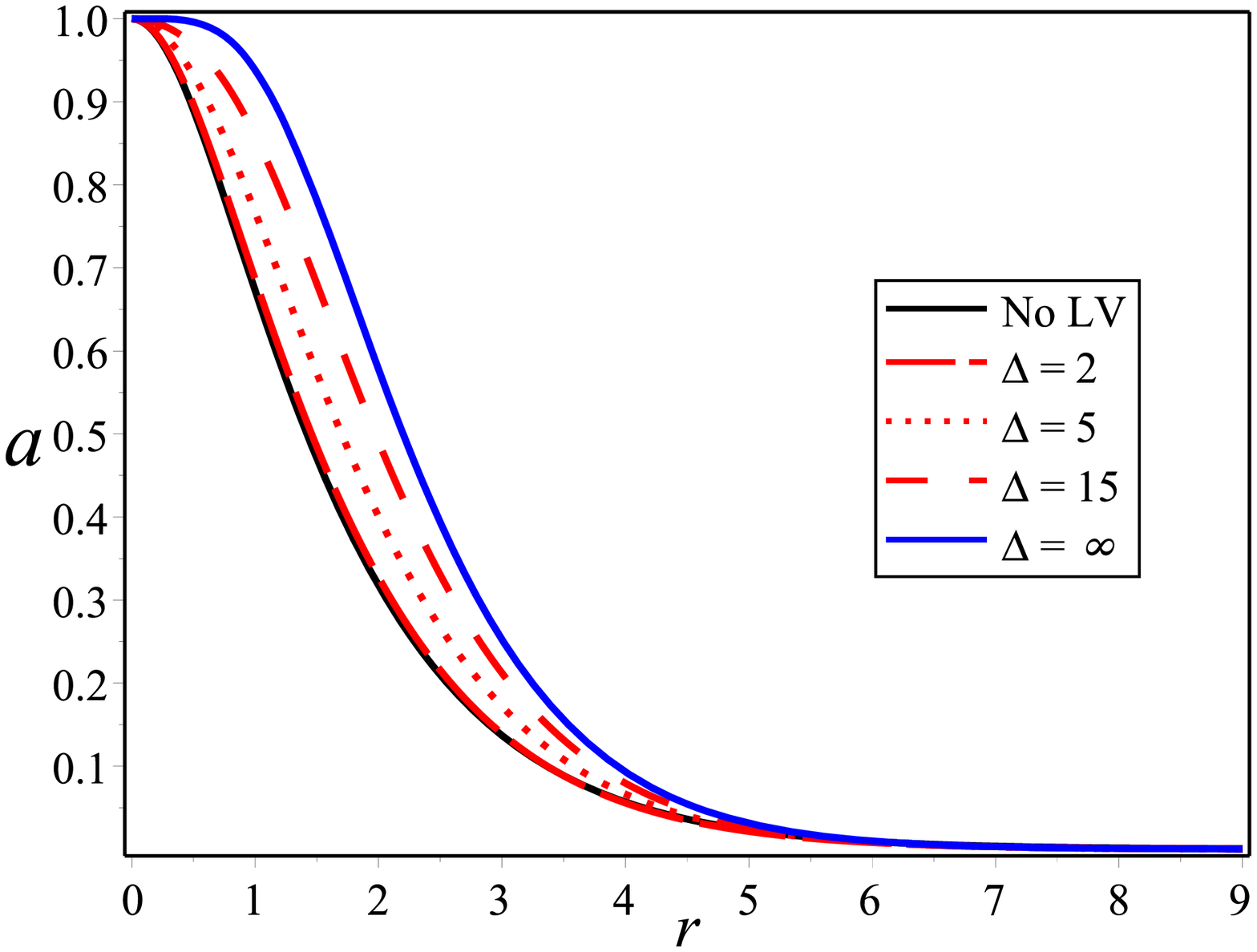}}
\caption{The profiles of the gauge field, $a(r),$ for winding number $n=1$.
The legends are the same as in Fig. \protect\ref{HIGGS}. }
\label{A}
\end{figure}

\begin{figure}[tbp]
\centering{\includegraphics[width=8.0cm]{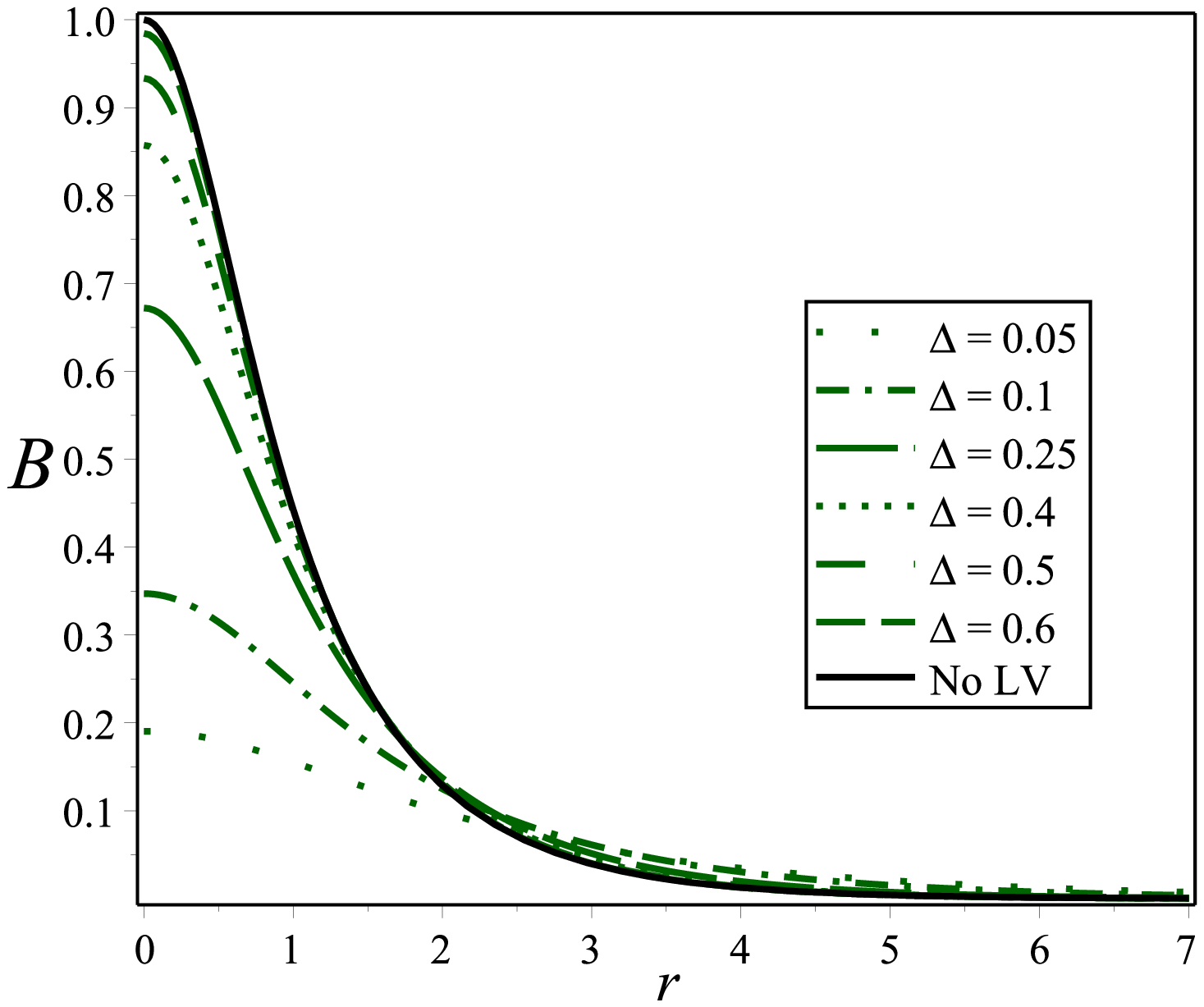} %
\includegraphics[width=8.0cm]{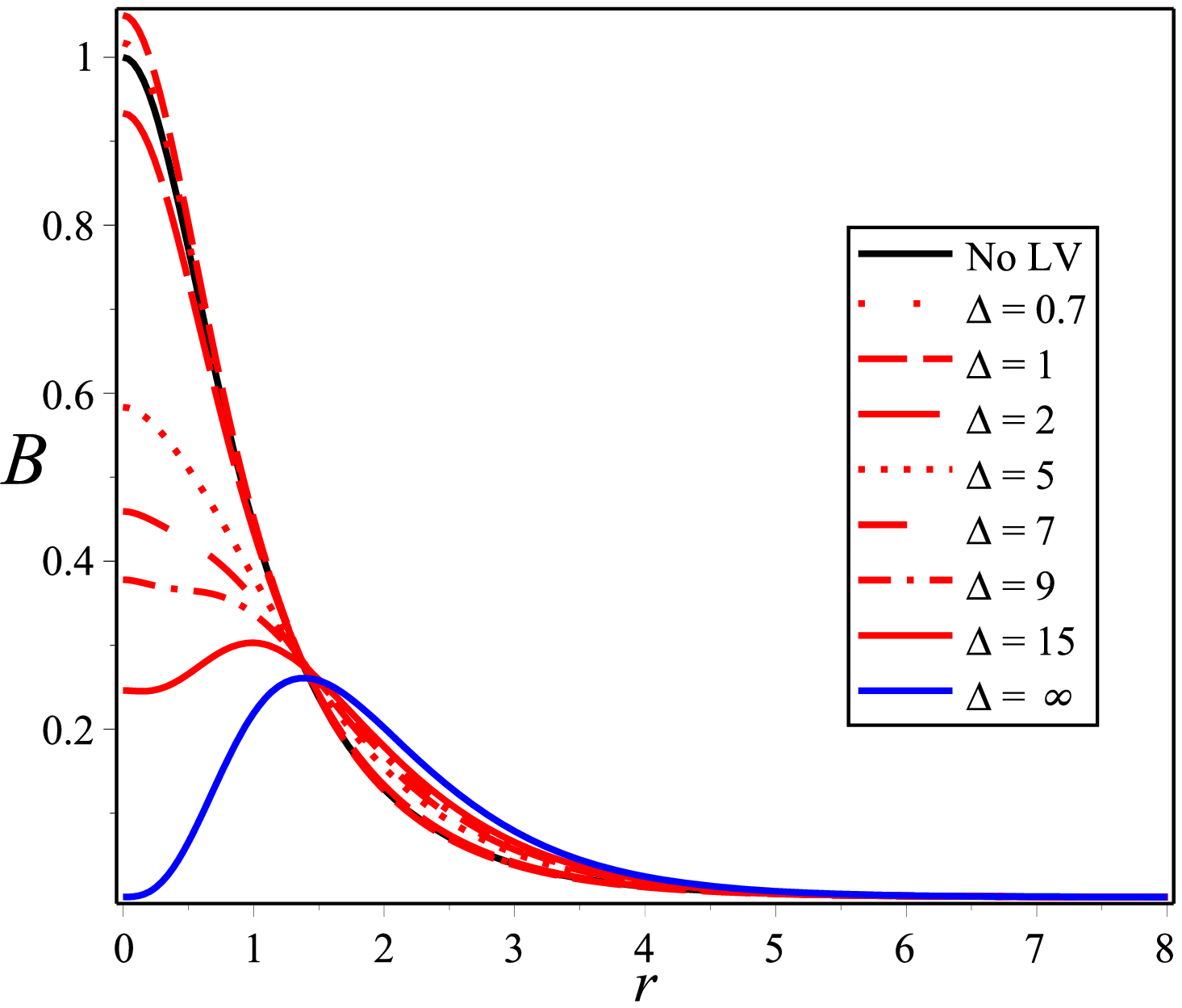}}
\caption{The profiles of the magnetic field, $B(r),$ for winding number $n=1$%
. The upper figure represents the profiles for $0<\Delta \leq 0.642$ and the
lower figure represents the profiles for $\Delta >0.642$. The solid black
line represents the profiles in the absence of Lorentz violation. The blue line
is the profile for $\Delta \rightarrow \infty $. }
\label{MF2}
\end{figure}

\begin{figure}[]
\centering\includegraphics[width=8.0cm]{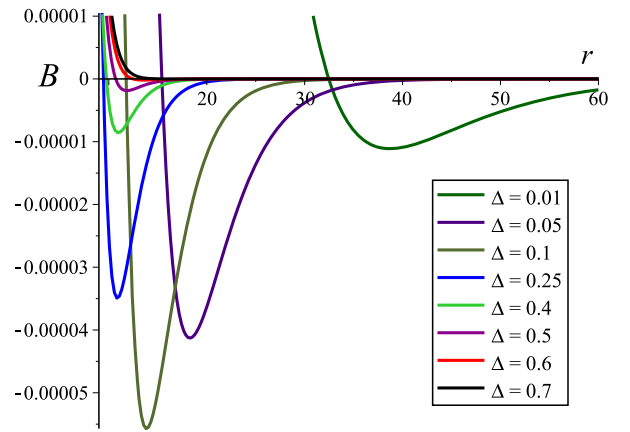}
\caption{Magnetic field inversion.}
\label{MF INV}
\end{figure}

\begin{figure}[]
\centering{\includegraphics[width=8.0cm]{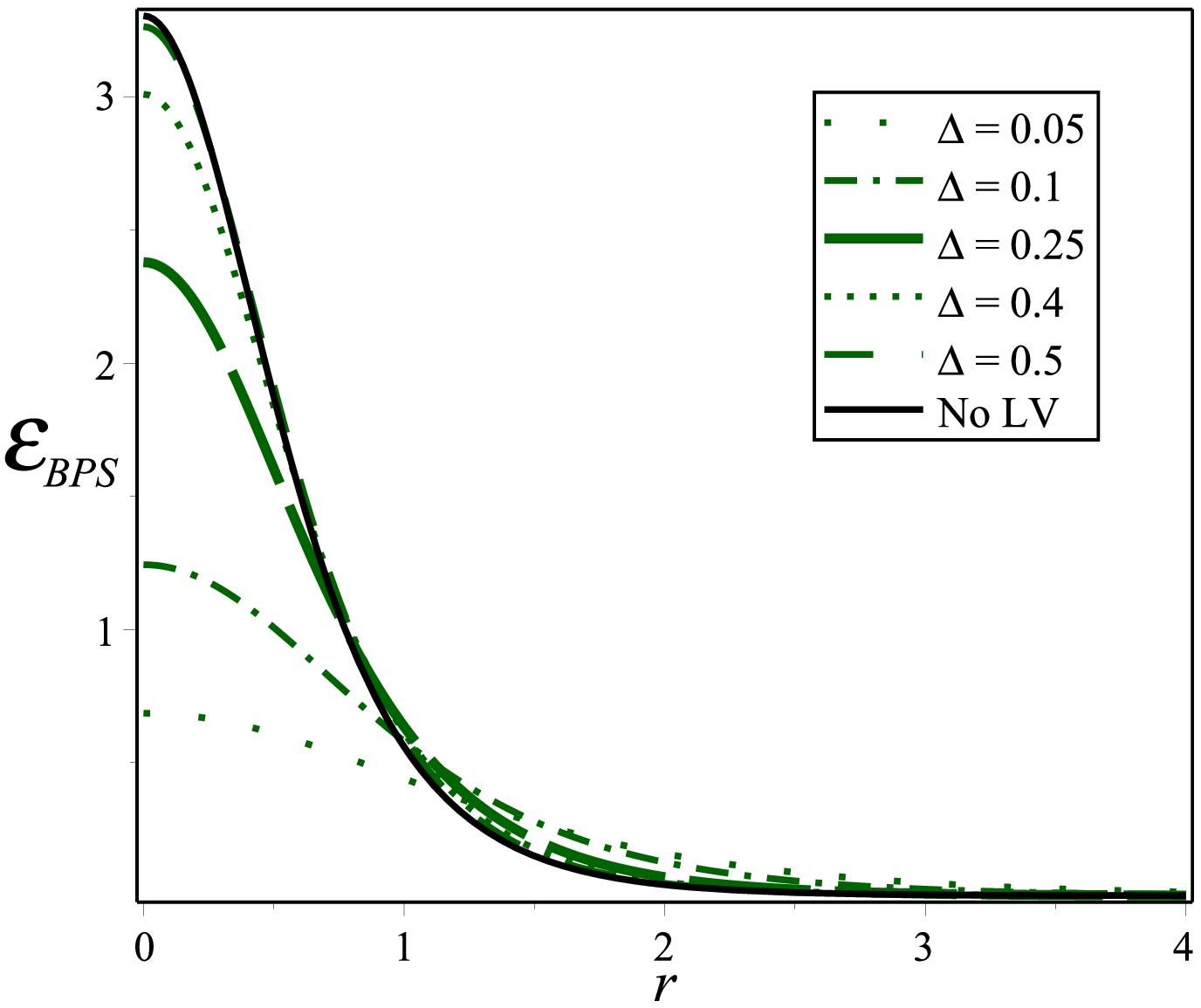} %
\includegraphics[width=8.0cm]{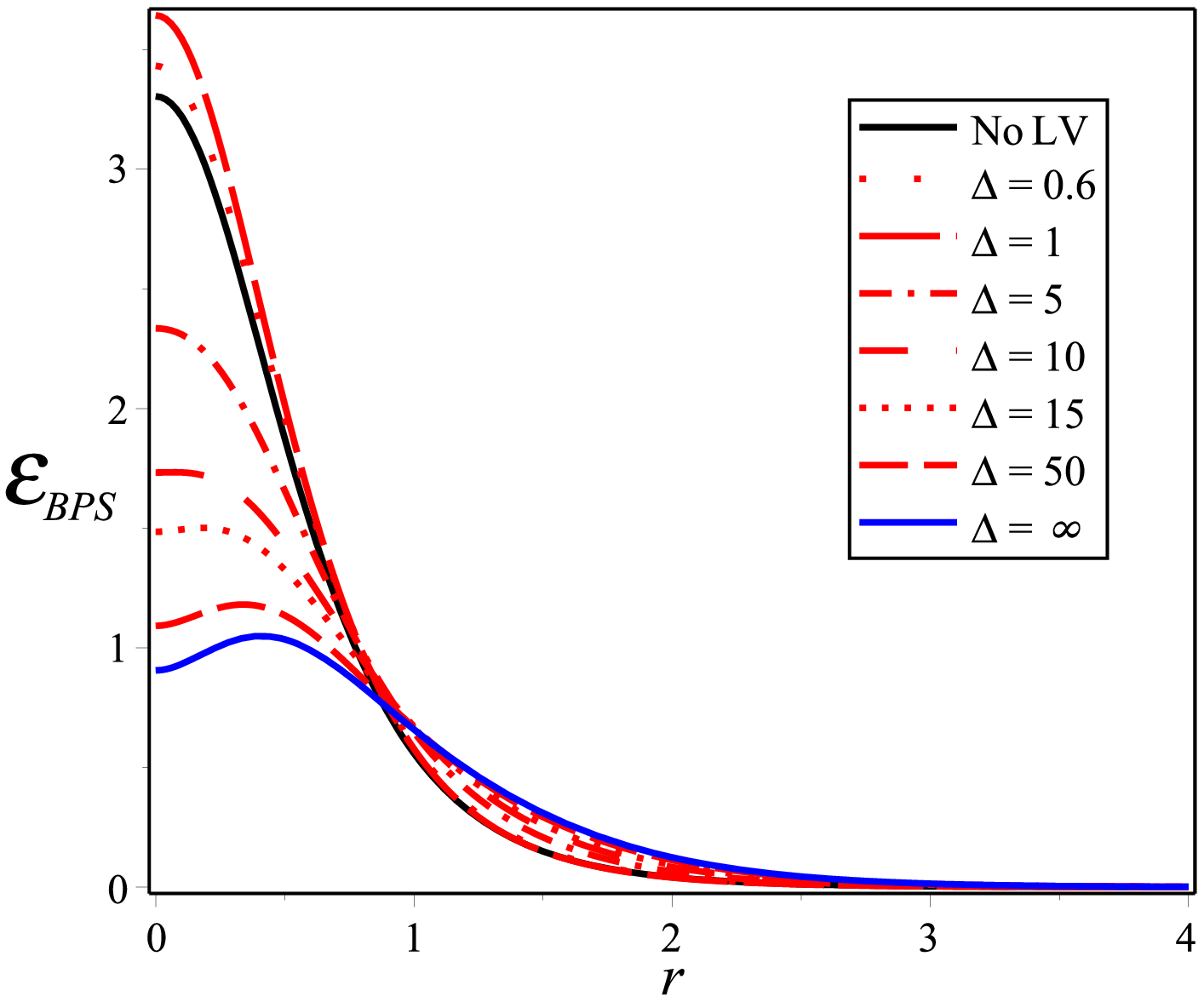}}
\caption{The profiles of the BPS energy density, $\protect\varepsilon %
_{bps}(r)$, for winding number $n=1$. The upper figure represents the
profiles for $0<\Delta \leq 0.5$ and the lower figure represents the profiles
for $\Delta \geq 0.6$. The solid black line represents the profiles in the
absence of Lorentz violation. The blue line is the profile for $\Delta
\rightarrow \infty $.}
\label{ENERGIA}
\end{figure}

We consider the case $\beta _{2}=0,$ with $\kappa _{ii}=0$, $\lambda _{r}=0$.
Then
\begin{equation}
\kappa _{0\theta }=\frac{\Delta -1}{2\sqrt{\Delta }}~,~\ {\Delta }>0\,.
\end{equation}%
In this way, the boundary conditions read
\begin{eqnarray}
g(0) &=&0,~a(0)=n,~A_{0}^{\prime }(0)=\frac{2\eta \left( \Delta -1\right)
\sqrt{\Delta }}{\left( 1+\Delta \right) ^{2}},  \notag \\[-0.15cm]
&&  \label{bcsxx} \\[-0.15cm]
g(\infty ) &=&\frac{\pi }{2},~a(\infty )=0,~A_{0}(\infty )=0.  \notag
\end{eqnarray}%
Hereafter, we consider\ $\Lambda =1$\ and $\eta =1.05$. As a consequence the
mass scale $m$\ is given by%
\begin{equation}
m=\sqrt{\frac{2.1\Delta }{1+{\Delta }}},  \label{betaMH}
\end{equation}%
taking\ the values, $0<m\leq \sqrt{2.1}$. We note that for $\Delta \ll 1$,
the defect reaches its asymptotic values slowly. But when $\Delta
\rightarrow 10/11$, the behavior is close to the profiles in the absence of
Lorentz violation, because $m\rightarrow 1$ (see the solid black lines in Figs. %
\ref{HIGGS}--\ref{ENERGIA}). On the other hand, for $\Delta \rightarrow
\infty $, the mass scale reaches its maximum value $m\rightarrow \sqrt{2.1}$
(see the solid blue lines in Figs. \ref{HIGGS}-\ref{ENERGIA}).

We have depicted the profiles obtained from numerical solutions of Eqs. (\ref%
{eqt2a})-(\ref{eqt2c}) under the boundary conditions (\ref{bcsxx}). Without
loss of generality, we have considered $n=1$, with the solutions being
compared with the profiles of the model in the absence of Lorentz violation
(the solid black lines).

Figures \ref{HIGGS} and \ref{A} show the profiles of the $\sigma $ field and
the gauge field, respectively. For $\Delta \ll 1,$ they are very spread and
reach their asymptotic value slowly. When $\Delta \rightarrow 1,$ they are
narrower and attain the vacuum state more rapidly. For $0<\Delta \leq 0.5,$
the profiles are limited by the one in the absence of Lorentz violation ($\Delta
=1$, solid black line). However, for $\Delta \geq 2$, the profiles become
wider but are limited by the width of the profile corresponding to $\Delta
\rightarrow \infty $ (solid blue line). Numerical analysis showed that the
profiles of the $\sigma $ field and the gauge field in the interval $%
0.5<\Delta <2$ are almost overlapped with the ones obtained in the absence of
Lorentz violation, which does not occur with the magnetic field and the BPS
energy density.

The magnetic field behavior is shown in Fig. \ref{MF2}. For the range $%
0<\Delta \leq 0.642$ (green lines), the profiles are lumps whose amplitudes
at the origin increases whenever $\Delta $ augments, reaching the value $%
B(0)=1$ for $\Delta =0.642$. For $0.642<\Delta <5$, the profiles are also
lumps centered at the origin. For $0.642<\Delta <1$ the amplitude increases
attaining its maximum value $B(0)=1.05$ when $\Delta =1$; on the other hand,
for $\Delta >1$ the profile amplitude decreases while $\Delta $ increases
continuously. For $5<\Delta <9$, the lumps present a deformation close to
the origin. An interesting fact is observed when $\Delta >9$: the deformed
lump begins to become a ringlike profile. This way, for large values of $%
\Delta $, the magnetic field approaches the CSH and MCSH ringlike profiles.
This is an interesting effect produced by the mixing of the parity-odd ($%
\kappa _{0\theta }$) gauge LV coefficient and the parity-even ($\Delta $)\
LV coefficient belonging to the $\sigma $ sector. Such a behavior of the
amplitude of the magnetic field at the origin can be verified by analyzing $%
B(0)$, which can be obtained directly from Eq. (\ref{BH_a0}):%
\begin{equation}
B(0)=\frac{4\eta \Delta }{\left( \Delta +1\right) ^{2}},
\end{equation}%
which behaves\ as $\Delta ^{-1}$\ for large values of $\Delta ,$ implying
that the magnetic field at the origin goes to zero when $\Delta \rightarrow \infty $%
. This result is valid for all values of winding number $n$.

Another remarkable feature of this model is the localized magnetic field
inversion. It takes place for $0\leq \Delta <0.7$, as is shown in
Fig. \ref{MF INV}, where a zoom was performed on the profiles with $0.01\leq
\Delta \leq 0.7$. One can clearly observe the localized magnetic field
inversion, which is more pronounced for values of $\Delta <0.6$. Around the
value $\Delta \simeq 0.6$ the inversion becomes negligible, ceasing for
larger values of $\Delta $.

The magnetic field inversion is a relevant feature of this model and can be
confirmed by means of an analytical analysis in the which one discusses the
behavior of the magnetic field for sufficiently large values of $r$ (i.e., $%
r\rightarrow \infty )$. Here, we analyze the case of positive winding
number. By considering $g\left( r\right) \rightarrow \pi /2$ when $%
r\rightarrow \infty $, the BPS equation (\ref{eqt2b}) is simplified as
\
\begin{equation}
B\left( r\right) \simeq -\frac{\kappa _{0\theta }}{1-\kappa _{ii}}%
A_{0}^{\prime }\left( r\right) .
\end{equation}%
The function $A_{0}^{\prime }\left( r\right) $ can be easily computed from
Eq. (\ref{bc-3a}),
\begin{equation}
A_{0}^{\prime }(r)\approx -\frac{\left( 1-\kappa _{ii}\right) m^{2}-\eta
\Lambda }{\Lambda \kappa _{0\theta }m}C_{_{\infty }}\left( m+\frac{1}{2r}%
\right) \frac{e^{-mr}}{\sqrt{r}}+\ldots,
\end{equation}%
so that, for large values of $r$, the magnetic field behaves as%
\begin{equation}
B\left( r\right) \simeq \frac{\left( 1-\kappa _{ii}\right) m^{2}-\eta
\Lambda }{\left( 1-\kappa _{ii}\right) \Lambda m}C_{_{\infty }}\left( m+%
\frac{1}{2r}\right) \frac{e^{-mr}}{\sqrt{r}}+\ldots
\end{equation}%
The only quantity that could be negative is~$\left( 1-\kappa _{ii}\right)
m^{2}-\eta \Lambda $. Our case, with $m$\ given by (\ref{mxx}), yields
\begin{equation}
\Delta \left( 1-\kappa _{ii}\right) <\left( 1+\lambda _{r}\right) ,
\end{equation}%
which, under the conditions $\kappa _{ii}=0=\lambda _{r}$ used in our
numerical analysis, provides a negative magnetic field for large values of $%
r $ when $\Delta <1$. This fact, associated with a positive $B(0)$, indicates
magnetic field inversion. This result is in complete agreement with the
profiles depicted in Fig. \ref{MF INV}. For $n<0$, the reciprocal situation
happens.

The magnetic field flipping finds applications in fractional vortices
occurring in superconductors described by the two-component Ginzburg-Landau
 model \cite{mff}.

Figure \ref{ENERGIA} presents the profiles for the BPS energy density, which
are very similar to the ones of the magnetic field. For $0<\Delta \lesssim
10 $, they are lumps centered at the origin whose amplitude increases when $%
\Delta $ falls in the range $0<\Delta \leq 1$, attaining its maximum value for $%
\Delta =1$. For $1<\Delta \lesssim 10$, the amplitude decreases while $%
\Delta $ increases. For $\Delta >10$, the profiles become ringlike
structures, with the behavior accentuated as $\Delta $ continuously grows. For $%
\Delta \rightarrow \infty $, the amplitude at the origin is $\sim 0.82$ (but
for $n>1$, such an amplitude is zero). Thus, for large values of $\Delta $,
the sigma model energy begins to behave as in the CSH and MCSH
models, a consequence of Lorentz violation. This can be understood by analyzing
the amplitude of the BPS energy density at the origin, which stems from Eq. (%
\ref{energy_BPS2}),
\begin{equation}
\varepsilon _{_{BPS}}(0)=\frac{4\Delta \eta ^{2}}{\left( \Delta +1\right)
^{2}}+\eta n^{2}\left( G_{n}\right) ^{2}r^{2\left( n-1\right) },
\end{equation}%
where it was verified that $G_{n}$\ is a finite quantity for any value of $%
\Delta $. Then, for large values of $\Delta $\ the amplitude is
\begin{equation}
\varepsilon _{_{BPS}}(0)=\frac{4\eta ^{2}}{\Delta }+\eta n^{2}\left(
G_{n}\right) ^{2}r^{2\left( n-1\right) },
\end{equation}%
which for $n=1$\ becomes $\varepsilon _{_{BPS}}(0)=\eta \left( G_{1}\right)
^{2}$, a finite quantity. On the other hand, for $n>1$, the amplitude goes
to zero as quickly as $\Delta ^{-1}~$does. In both cases, the numerical
result is verified by the analytical analysis.

\section{Remarks and conclusions}

We have examined a gauged $O(3)$\ $\sigma $ model modified by
Lorentz-violating terms in the non-Abelian scalar and electromagnetic
sectors, demonstrating the existence of topological self-dual
configurations. In the standard gauged $O(3)$ $\sigma $ model only
purely magnetic self-dual configurations exists. The introduction of LV terms
allows the existence of altered purely magnetic self-dual solutions and
magnetic self-dual configurations carrying an electric field. Specifically, the
purely magnetic self-dual configurations take place when the parity-odd,
$CPT$-even coefficients, $\left( k_{\phi \phi }\right) _{0i}=0,$ $\kappa
_{0i}=0,$ are null. For $\kappa _{0i}\neq 0,$ the self-dual configurations
also carry an electric field but a null total electric charge. Implementing the
BPS procedure, the total energy of the self-dual configurations in both
cases was evaluated, revealing itself to be proportional to the topological charge
of the model and to the LV coefficients introduced in the $\sigma $ sector.
It was noticed that, while the purely magnetic configurations were
quantitatively altered in they widths by the LV terms, the
charged configurations may undergo sensitive qualitative modification by
the same terms, approaching the magnetic and energy behavior of the CSH and
MCSH models. Furthermore, also reported was the remarkable possibility of
magnetic flux reversion, which finds an application in some condensed matter
systems. Therefore, we stress that the Lorentz violation significantly enriches the
space of self-dual configurations found in the sigma model of Ref. \cite%
{mukherjee2}.

\acknowledgments

R. C. and M. M. F. Jr are grateful to CNPq (Conselho
Nacional de Desenvolvimento Cient\'{i}fico e Tecnol\'{o}gico),
CAPES (Coordena\c{c}\~{a}o de Aperfei\c{c}oamento de Pessoal de
N\'{i}vel Superior) and FAPEMA (Funda\c{c}\~{a}o de Amparo \`a
Pesquisa e ao Desenvolvimento Cient\'{i}fico e Tecnol\'{o}gico do
Maranh\~{a}o), brazilian funding agencies for development of
science; C. F. acknowledges CAPES for the invaluable
financial support.


\begin{thebibliography}{99}
\bibitem{ano} A. A. Abrikosov, Zh. Eksp. Teor. Fiz. {\bf 32}, 1442
(1957) [Sov. Phys. JETP {\bf 5}, 1174 (1957)];

\bibitem{ginz} V. L. Ginzburg and L. D. Landau, Zh. Eksp. Teor. Fiz. {\bf 20},
1064 (1950); in {\it Collected Papers of L. D. Landau}, edited by D. Ter Haar (Pergamon Press, Oxford, 1965), p. 546.

\bibitem{nielsen} H. Nielsen and P. Olesen, Nucl. Phys. \textbf{B 61}, 45
(1973).

\bibitem{jackiw1} S. Deser, R. Jackiw, and S. Templeton, Ann. Phys. (N.Y.)
\textbf{140}, 372 (1982); G. V. Dunne, Aspects of Chern-Simons Theory, arXiv:hep-th/9902115.

\bibitem{jackiw2} R. Jackiw and E. J. Weinberg, Phys. Rev. Lett. \textbf{64}%
, 2234 (1990); R. Jackiw, K. Lee, and E. J. Weinberg, Phys. Rev. D \textbf{42}%
, 3488 (1990); J. Hong, Y. Kim, and P. Y. Pac, Phys. Rev. Lett. \textbf{64},
2230 (1990); G. V. Dunne, {\it Self-Dual Chern-Simons Theories} (Springer,
Heidelberg, 1995).

\bibitem{lee} C.k. Lee, K. M. Lee, and H. Min, Phys. Lett. B \textbf{252}, 79
(1990).

\bibitem{polyakov} A. A. Belavin and A. M. Polyakov, JETP Lett. \textbf{22},
245 (1975).

\bibitem{CMP} R. Rajaraman, {\it Solitons and Instantons} (North-Holland,
Amsterdam 1982); W. J. Zakrzewski, {\it Low Dimensional Sigma Models} (Hilger,
Bristol, England, 1989).

\bibitem{bog} E. B. Bogomol'nyi, Sov. J. Nucl. Phys. \textbf{24}, 449 (1976);
M. K. Prasad and C. M. Sommerfield, Phys. Rev. Lett. \textbf{35}, 760 (1975).

\bibitem{zakrzewski} R. A. Leese, M. Peyrard, and W. J. Zakrzewski,
Nonlinearity \textbf{3}, 387 (1990).

\bibitem{leese1} M. Peyrard, B. M. A. G. Piette, and W. J. Zakrzewski,
Nonlinearity \textbf{5}, 563 (1992).

\bibitem{leese2} R. A. Leese, Nucl. Phys. \textbf{B344}, 33 (1990); Nucl.
Phys. \textbf{B366}, 283 (1991).

\bibitem{schroers} B. J. Schroers, Phys. Lett. B \textbf{356}, 291 (1995).

\bibitem{ghosh} P. K. Ghosh and S. K. Ghosh, Phys. Lett. B \textbf{366}, 199
(1996).

\bibitem{mukherjee1} P. Mukherjee, Phys. Lett. B \textbf{403}, 70 (1997).

\bibitem{mukherjee2} P. Mukherjee, Phys. Rev. D \textbf{58}, 105025 (1998).

\bibitem{almeida} F. S. A. Cavalcante, M. S. Cunha, and C. A. S. Almeida, Phys.
Lett. B \textbf{475}, 315 (2000); M. S. Cunha, R. R. Landim and C. A. S. Almeida,
Phys. Rev. D \textbf{74}, 067701 (2006).

\bibitem{bazeia} D. Bazeia, E. da Hora, R. Menezes, H. P. de Oliveira, and C.
dos Santos, Phys. Rev. D \textbf{81}, 125016 (2010); D. Bazeia, E. da Hora,
C. dos Santos, and R. Menezes, Phys. Rev. D \textbf{81}, 125014 (2010); D.
Bazeia, E. da Hora, and R. Menezes, Phys. Rev. D \textbf{85}, 045005 (2012).

\bibitem{kost97} D. Colladay and V. A. Kostelecky, Phys. Rev. D \textbf{55},
6760 (1997); \textbf{58}, 116002 (1998).

\bibitem{coleman} S. R. Coleman and S. L. Glashow, Phys. Rev. D \textbf{59},
116008 (1999).

\bibitem{Vertex} F. R. Klinkhamer and M. Schreck, Nucl. Phys. \textbf{B848},
90 (2011); M. Schreck, Phys. Rev. D \textbf{\ 86}, 065038 (2012); M. A.
Hohensee, R. Lehnert, D. F. Phillips, and R. L. Walsworth, Phys. Rev. D
\textbf{80}, 036010 (2009); A. Moyotl, H. Novales-S\'{a}nchez, J. J.
Toscano, and E. S. Tututi, Int. J. Mod. Phys. A \textbf{29}, 1450039 (2014);
\textbf{29}, 1450107 (2014); M. Cambiaso, R. Lehnert, and R.
Potting, Phys. Rev. D \textbf{90}, 065003 (2014); R. Bufalo, Int. J. Mod.
Phys. A \textbf{29}, 1450112 (2014).


\bibitem{bm} V. A. Kostelecky, Phys. Rev. D\textbf{\ 69}, 105009 (2004); R.
Bluhm and V. A. Kosteleck\'{y}, Phys. Rev. D \textbf{71}, 065008 (2005); Q.
G. Bailey and V. A. Kosteleck\'{y}, Phys. Rev. D \textbf{74}, 045001 (2006); C.
Hernaski, Phys. Rev. D \textbf{90}, 124036 (2014); R. V. Maluf, J. E. G.
Silva, and C. A. S. Almeida, Phys. Lett. B \textbf{749}, 304 (2015).

\bibitem{fermion} V. A. Kostelecky and C. D. Lane, J. Math. Phys. (N.Y.) \textbf{40}%
, 6245 (1999); R. Lehnert, J. Math. Phys. (N.Y.) \textbf{45}, 3399 (2004);\ D.
Colladay and V. A. Kostelecky, Phys. Lett. B \textbf{511}, 209 (2001); T.
Mariz, J. R. Nascimento, and A. Yu. Petrov, Phys. Rev. D \textbf{85}, 125003
(2012); G. Gazzola, H. G. Fargnoli, A. P. Baeta Scarpelli, M. Sampaio, and
M. C. Nemes, J. Phys. G \textbf{39}, 035002 (2012); A. P. Baeta Scarpelli,
M. Sampaio, M. C. Nemes, and B. Hiller, Eur. Phys. J. C \textbf{56}, 571
(2008); F. A. Brito, L. S. Grigorio, M. S. Guimaraes, E. Passos, and C. Wotzasek,
Phys.Rev. D \textbf{78}, 125023 (2008); F. A. Brito, E. Passos, and P. V. Santos,
Europhys. Lett. \textbf{95}, 51001 (2011); C. F. Farias, A. C. Lehum, J. R.
Nascimento, and A. Yu. Petrov, Phys. Rev. D \textbf{86}, 065035 (2012); J. R.
Nascimento, A. Yu. Petrov, C. Wotzasek, and C. A. D. Zarro, Phys. Rev. D \textbf{%
89}, 065030 (2014); O. M. Del Cima, J. M. Fonseca, D. H.T. Franco, O.
Piguet, Phys. Lett. B \textbf{688}, 258 (2010). R. V. Maluf, J. E. G. Silva, W. T. Cruz, and C. A. S. Almeida,
Phys. Lett. B \textbf{738}, 341 (2014).

\bibitem{new} V. A. Kostelecky, C. D. Lane, A. G. M. Pickering, Phys. Rev. D \textbf{65}, 056006
(2002); C. D. Carone, M. Sher, M. Vanderhaeghen, Phys. Rev. D \textbf{74}, 077901
(2006); W. F. Chen, G. Kunstatter, Phys. Rev. D \textbf{62}, 105029 (2000); O. M. Del Cima,
D. H. T. Franco, A. H. Gomes, J. M. Fonseca, O. Piguet, Phys. Rev. D
\textbf{85}, 065023 (2012); T. R. S. Santos, R. F. Sobreiro, Phys. Rev. D \textbf{91}, 025008
(2015).

\bibitem{carroll} S. M. Carroll, G. B. Field, and R. Jackiw, Phys. Rev. D
\textbf{41}, 1231 (1990).

\bibitem{mewes1} V. A. Kostelecky and M. Mewes, Phys. Rev. Lett. \textbf{87}%
, 251304 (2001); Phys. Rev. D \textbf{66}, 056005 (2002);
Phys. Rev. Lett. \textbf{97}, 140401 (2006).

\bibitem{altschul} B. Altschul, Nucl. Phys. \textbf{B796}, 262 (2008);
Phys. Rev. Lett. \textbf{98}, 041603 (2007); C. Kaufhold and F. R.
Klinkhamer, Phys. Rev. D \textbf{76}, 025024 (2007).

\bibitem{klinkhamer1} F. R. Klinkhamer and M. Risse, Phys. Rev. D \textbf{77}%
, 016002 (2008); \textbf{77}, 117901 (2008); F. R. Klinkhamer and M.
Schreck, Phys. Rev. D \textbf{78}, 085026 (2008).

\bibitem{HD} M. Cambiaso, R. Lehnert, and R. Potting, Phys. Rev. D \textbf{85%
}, 085023 (2012); B. Agostini, F. A. Barone, F. E. Barone, P. Gaete, and J.
A. Helay\"{e}l-Neto, Phys. Lett. B \textbf{708}, 212 (2012); L. Campanelli,
Phys. Rev. D \textbf{90}, 105014 (2014); R. Bufalo, B. M. Pimentel, and D. E.
Soto, Phys. Rev. D \textbf{90}, 085012 (2014).

\bibitem{Reyes} R. C. Myers and M. Pospelov, Phys. Rev. Lett. \textbf{90},
211601 (2003); C. M. Reyes, L. F. Urrutia, and J. D. Vergara, Phys. Rev. D
\textbf{78}, 125011 (2008); Phys. Lett. B \textbf{675}, 336 (2009); C. M.
Reyes, Phys. Rev. D \textbf{82}, 125036 (2010); \textbf{80},
105008 (2009); \textbf{87}, 125028 (2013); C. M. Reyes, S.
Ossandon, and C. Reyes, Phys. Lett. B \textbf{746}, 190 (2015).

\bibitem{anacleto} M. A. Anacleto, F. A. Brito, and E. Passos, Phys. Rev. D
\textbf{86}, 125015 (2012); M. A. Anacleto, Phys. Rev. D \textbf{92}, 085035
(2015); E. O. Silva and F. M. Andrade, Europhys. Lett. \textbf{101}, 51005
(2013); F. M. Andrade, E. O. Silva, T. Prud\^{e}ncio and C. Filgueiras, J.
Phys. G \textbf{40} 075007 (2013).

\bibitem{bakke} K. Bakke and H. Belich, J. Phys. G \textbf{39}, 085001
(2012); K. Bakke, H. Belich, and E. O. Silva, J. Math. Phys. (N.Y.) \textbf{52},
063505 (2011); \ J. Phys. G \textbf{39}, 055004 (2012); K. Bakke and H.
Belich, J. Phys. G \textbf{39}, 085001 (2012); Ann. Phys. (N.Y.) \textbf{333}, 272 (2013);
A. G. de Lima, H. Belich, and K. Bakke, Ann. Phys. (Berlin) \textbf{526}, 514 (2014);
Eur. Phys. J. Plus \textbf{128}, 154 (2013).

\bibitem{scalar} M. N. Barreto, D. Bazeia, and R. Menezes, Phys. Rev. D \textbf{%
73}, 065015 (2006); A. de Souza Dutra, M. Hott, and F. A.Barone, Phys. Rev.
D \textbf{74}, 085030 (2006); D. Bazeia, M. M. Ferreira, Jr., A. R. Gomes, R.
Menezes, Physica (Amsterdam) \textbf{239D}, 942 (2010); A. de Souza Dutra and R. A. C.
Correa, Phys. Rev. D \textbf{83}, 105007 (2011); R. A. C. Correa, R. da Rocha, A. de Souza Dutra, Ann.
Phys. \textbf{359}, 198~(2015).

\bibitem{barraz} N. M. Barraz, Jr., J. M. Fonseca, W. A. Moura-Melo, and J. A.
Helayel-Neto, Phys.Rev. D \textbf{76}, 027701 (2007); A. P. Baeta Scarpelli
and J. A. Helayel-Neto, Phys.Rev. D \textbf{73}, 105020 (2006).

\bibitem{seifert} M. D. Seifert, Phys. Rev. Lett. \textbf{105}, 201601
(2010); Phys. Rev. D \textbf{82}, 125015 (2010).

\bibitem{correa1} A. de Souza Dutra and R. A. C. Correa, Adv. High Energy Phys.
\textbf{2015}, 673716 (2015).

\bibitem{correa3} R. A. C. Correa, Roldao da Rocha, A. de Souza Dutra, Phys.
Rev. D \textbf{91}, 125021 (2015).

\bibitem{miller} C. Miller, R. Casana, M. M. Ferreira, Jr., and E. da Hora,
Phys.Rev. D \textbf{86}, 065011 (2012).

\bibitem{casana1} R. Casana, M. Ferreira, Jr., E. da Hora, and C. Miller,
Phys. Lett. B \textbf{718}, 620 (2012).

\bibitem{sourrouille} L. Sourrouille, Phys. Rev. D \textbf{89}, 087702
(2014); R. Casana and L. Sourrouille, Phys. Lett. B \textbf{726}, 488 (2013).

\bibitem{hott} C. H. Coronado Villalobos, J. M. Hoff da Silva, M. B. Hott, and H.
Belich, Eur. Phys. J. C \textbf{74}, 2799 (2014).

\bibitem{belich} H. Belich, F. J. L. Leal, H. L. C. Louzada, and M. T. D. Orlando,
Phys. Rev. D \textbf{86}, 125037 (2012).

\bibitem{Guillermo} R. Casana and G. Lazar, Phys. Rev. D \textbf{90}, 065007
(2014).

\bibitem{sigmaMCSH} K. Kimm, K. Lee, and T. Lee, Phys. Rev. D \textbf{53} 4436
(1996); J. Han and H.-S. Nam, Lett. Math. Phys. \textbf{73}, 17 (2005).

\bibitem{bolog} S. Bolognesi and S. B. Gudnason, Nucl. Phys. \textbf{B805}, 104
(2008).

\bibitem{susycsh} B.-H. Lee, C.-k. Lee, and H. Min, Phys. Rev. D
\textbf{45}, 4588 (1992).

\bibitem{WittenOlive} E. Witten and D. Olive, Phys. Lett. \textbf{78B}, 97
(1978).

\bibitem{mff} E. Babaev, J. J\"{a}ykk\"{a}, and M. Speight, Phys. Rev. Lett.
\textbf{103}, 237002 (2009).
\end{thebibliography}
\end{document}